\documentclass[onecolumn,aps,prx,longbibliography,superscriptaddress,floatfix,10pt]{revtex4-2}

\usepackage[most]{tcolorbox}

\usepackage{amsmath,amssymb}
\usepackage{amsthm}
\usepackage{physics}
\usepackage{tikz}
\usepackage{graphicx}
\usepackage{float}
\usepackage[caption=false]{subfig}
\graphicspath{{figures/}}
\usepackage{dcolumn}
\usepackage{enumitem}
\usepackage[normalem]{ulem}
\usepackage{float}
\usepackage{bm,dsfont}
\usepackage{multirow}
\usepackage{simpler-wick}
\usepackage{cuted}
\usepackage{color}
\usepackage{hyperref}
\hypersetup{
	colorlinks = true,
	linkcolor = [rgb]{0.70,0.13,0.13},
	citecolor = [rgb]{0.13,0.55,0.13},
	urlcolor = [rgb]{0.25, 0.41, 0.88}}
\usepackage[capitalize]{cleveref}
\newtcolorbox[use counter=alg,crefname={algorithm}{algorithms},Crefname={Algorithm}{Algorithms}]{alg}[2][]{float=h,colback=yellow!5!white,colframe=yellow!50!black,  colbacktitle=yellow!75!black,fonttitle=\bfseries,
title=Algorithm~\thetcbcounter: #2,#1}

\newcommand{\id}{\mathds{1}}

\newcommand{\dsE}{\mathbb{E}}

\newcommand{\sC}{\mathcal{C}
}
\newcommand{\Cl}{{\mathcal{C}\ell}}
\newcommand{\scD}{\mathcal{D}}
\newcommand{\scE}{\mathcal{E}}

\newcommand{\scM}{\mathcal{M}}

\newcommand{\scO}{\mathcal{O}}

\DeclareMathOperator{\swap}{SWAP}
\DeclareMathOperator{\diag}{diag}

\let\Cl\relax
\DeclareMathOperator{\Cl}{Cl}

\newcommand{\darkcirc}{\tikz \fill (0,0) circle (0.7ex);}
\newcommand{\whitecirc}{\tikz \draw (0,0) circle (0.7ex);}

\newcommand{\eqs}[1]{\begin{equation}\begin{split}#1\end{split}\end{equation}}
\newcommand{\rket}[1]{|#1)}
\newcommand{\rbra}[1]{(#1|}

\newtheorem{theorem}{Theorem}
\newtheorem{lemma}[theorem]{Lemma}
\newtheorem*{lemma*}{Lemma}

\usepackage[mathlines]{lineno}

\begin{document}

\title{Demonstration of Robust and Efficient Quantum Property Learning with \\ Shallow Shadows}
\author{Hong-Ye Hu}
\thanks{Equal contributions}
\affiliation{Department of Physics, Harvard University, 17 Oxford Street, Cambridge, MA 02138, USA}
\author{Andi Gu}
\thanks{Equal contributions}
\affiliation{Department of Physics, Harvard University, 17 Oxford Street, Cambridge, MA 02138, USA}
\author{Swarnadeep Majumder}
\thanks{Equal contributions}
\affiliation{IBM Quantum, IBM T.J. Watson Research Center, Yorktown Heights, NY 10598, USA}
\author{Hang Ren}
\affiliation{Berkeley Center for Quantum Information and Computation, Berkeley, California 94720, USA}
\author{Yipei Zhang}
\affiliation{Berkeley Center for Quantum Information and Computation, Berkeley, California 94720, USA}
\author{Derek~S. Wang}
\affiliation{IBM Quantum, IBM T.J. Watson Research Center, Yorktown Heights, NY 10598, USA}
\author{Yi-Zhuang You}
\email{yzyou@physics.ucsd.edu}
\affiliation{Department of Physics, University of California San Diego, La Jolla, CA 92093, USA}
\author{Zlatko Minev}
\email{zlatko.minev@ibm.com}
\affiliation{IBM Quantum, IBM T.J. Watson Research Center, Yorktown Heights, NY 10598, USA}
\author{Susanne F. Yelin}
\email{syelin@g.harvard.edu}
\affiliation{Department of Physics, Harvard University, 17 Oxford Street, Cambridge, MA 02138, USA}
\author{Alireza Seif}
\email{alireza.seif@ibm.com}
\affiliation{IBM Quantum, IBM T.J. Watson Research Center, Yorktown Heights, NY 10598, USA}

\date{\today}
\begin{abstract}
Extracting information efficiently from quantum systems is a major component of quantum information processing tasks. Randomized measurements, or classical shadows, enable predicting many properties of arbitrary quantum states using few measurements. While random single-qubit measurements are experimentally friendly and suitable for learning low-weight Pauli observables, they perform poorly for nonlocal observables. Prepending a shallow random quantum circuit before measurements maintains this experimental friendliness, but also has favorable sample complexities for observables beyond low-weight Paulis, including high-weight Paulis and global low-rank properties such as fidelity. However, in realistic scenarios, quantum noise accumulated with each additional layer of the shallow circuit biases the results. To address these challenges, we propose the \emph{robust shallow shadows protocol}. Our protocol uses Bayesian inference to learn the experimentally relevant noise model and mitigate it in postprocessing. This mitigation introduces a bias-variance trade-off: correcting for noise-induced bias comes at the cost of a larger estimator variance. Despite this increased variance, as we demonstrate on a superconducting quantum processor, our protocol correctly recovers state properties such as expectation values, fidelity, and entanglement entropy, while maintaining a lower sample complexity compared to the random single qubit measurement scheme. We also theoretically analyze the effects of noise on sample complexity and show how the optimal choice of the shallow shadow depth varies with noise strength. This combined theoretical and experimental analysis positions the robust shallow shadow protocol as a scalable, robust, and sample-efficient protocol for characterizing quantum states on current quantum computing platforms.

\end{abstract}

\maketitle

\section{Introduction}

Classical shadow tomography \cite{HuangKeungPreskill} has emerged as a useful technique for efficiently characterizing quantum states with few measurements. This method leverages randomized measurements~\cite{Notarnicola2023A2112.11046, Elben2023T2203.11374} to construct a classical approximation or ``shadow'' of a quantum state, enabling the estimation of various state properties without the need for costly protocols such as full state tomography~\cite{Haah2015S1508.01797, ODonnell2015E1508.01907, Flammia2012Q1205.2300}. This method is particularly attractive because it allows experimentalists to `measure first and ask questions later'~\cite{Elben2023T2203.11374}: the same dataset can be used multiple times to learn a wide class of state properties. As such, classical shadow tomography and its variations~\cite{Chen2021R2011.09636, Enshan-Koh2020C2011.11580, Huang2021E2103.07510, Zhao2021F2010.16094, Hu2022H2102.10132, Hu2023C2107.04817, Bu2024C2202.03272, Akhtar2023S2209.02093, Bertoni2022S2209.12924, Nguyen2022O, Zhou2023P2212.11068, Zhou2023E2309.01258, Wu2023E2310.12726,PhysRevLett.120.050406,doi:10.1126/science.aau4963,2021arXiv210505992A,2022arXiv220808416Z,PhysRevResearch.3.033155,PhysRevLett.131.240602,Zhou2023performanceanalysis,2023CMaPh.404..629W,PhysRevX.13.011049,PhysRevLett.131.160601,2023arXiv230912933D,2023arXiv230910745I,Van-Kirk2022H2212.06084,2023arXiv231100695L,PhysRevLett.125.200501,2024arXiv240116922F,2024arXiv240118071F,2022arXiv221109835A} have found applications in a broad spectrum of quantum information tasks, including state verification \cite{Lukens2021A2012.08997, Morris2021Q2109.03860}, device benchmarking \cite{Levy2024C2110.02965, Kunjummen2023S2110.03629, Helsen2023S2110.13178,CrossPlatform,PRXQuantum.2.010102,PhysRevLett.124.010504}, Hamiltonian learning \cite{Hadfield2020M2006.15788, McNulty2023E2206.08912, Dutt2023P2312.07497,EntanglementHamiltonian,PhysRevLett.127.170501}, error mitigation \cite{Hu2022L2203.07263,Seif2023S2203.07309,2023arXiv230504956J,AndrewZhao}, and quantum machine learning \cite{2023arXiv230600061J,2023arXiv230614838Z,doi:10.1126/science.abk3333,2023MLS&T...4a5005H}.

However, classical shadows are not a panacea: a poor choice of randomized measurement scheme can result in poor performance. For instance, although random Pauli measurements are experimentally friendly and perform well for recovering low-weight Pauli observables, they are known to require high sample complexities for predicting nonlocal Pauli observables and low-rank global observables such as fidelity~\cite{HuangKeungPreskill}. On the other hand, schemes that use fully global random twirling (i.e., global random Clifford unitaries) are well-suited for low-rank global observables, yet they are experimentally infeasible due to the long circuit depths required to implement a global twirling unitary. These limitations have motivated the exploration of alternative randomized measurement schemes that maintain experimental feasibility, yet achieve improved sample complexity scaling on a broader class of observables. To list just a few, these alternative schemes include Hamiltonian-driven systems \cite{Hu2022H2102.10132,PhysRevX.13.011049,2023arXiv231100695L}, locally-scrambled quantum dynamics \cite{Hu2023C2107.04817,Bu2024C2202.03272,Zhou2023E2309.01258}, and shallow quantum circuits \cite{Akhtar2023S2209.02093,Bertoni2022S2209.12924}. Among these, measurement schemes using random finite-depth quantum circuits, referred to as shallow shadows, have been shown to have considerably lower sample complexities for predicting nonlocal and low-rank observables. 


While classical shadow tomography has been successfully demonstrated experimentally using random Pauli measurements \cite{doi:10.1126/science.abn7293,CrossPlatform,Struchalin2020E2008.05234, Zhang2021E2106.10190,2023arXiv230716882V}, the shallow shadows protocol --- which offers theoretical advantages for certain observables --- has never been experimentally validated on real quantum devices. This is important because these devices are noisy: until now, it has remained unclear whether, and to what degree, the benefits of shallow shadows can persist even under the presence of noise. Indeed, a blind application of existing theoretical protocols, without accounting for the effects of noise, will produce biased results in real experiments. 

In this work, we introduce the robust shallow shadows protocol, which is designed to be robust against the inherent noise in quantum systems. More specifically, our protocol aims to accurately predict a broad spectrum of quantum state properties from a single set of randomized measurements conducted on noisy, shallow quantum circuits. Our research advances the theoretical understanding of the shallow shadows protocol in three key ways. First, we introduce a robust shallow shadows protocol that produces unbiased classical shadows, enabling the efficient and noise-resilient prediction of many quantum state properties. This protocol requires only a simple calibration experiment and minimal assumptions. We also bound the sample complexity of the calibration and demonstrate that our method generalizes the robust classical shadow approach based on random Pauli or Clifford measurements. Second, we prove that incorporating a stochastic Pauli noise model in post-processing effectively captures time-independent experimental noise, and by combining this with Bayesian learning, we reduce the calibration sample overhead. Third, we explore the bias-variance tradeoff, showing that noise prevents the system from reaching the optimal noiseless circuit depth identified in Ref.~\cite{Ippoliti_2023}. We quantify how this optimal depth decreases with increasing noise strength.

\begin{figure}[htbp]
    \centering
    \includegraphics[width=\textwidth]{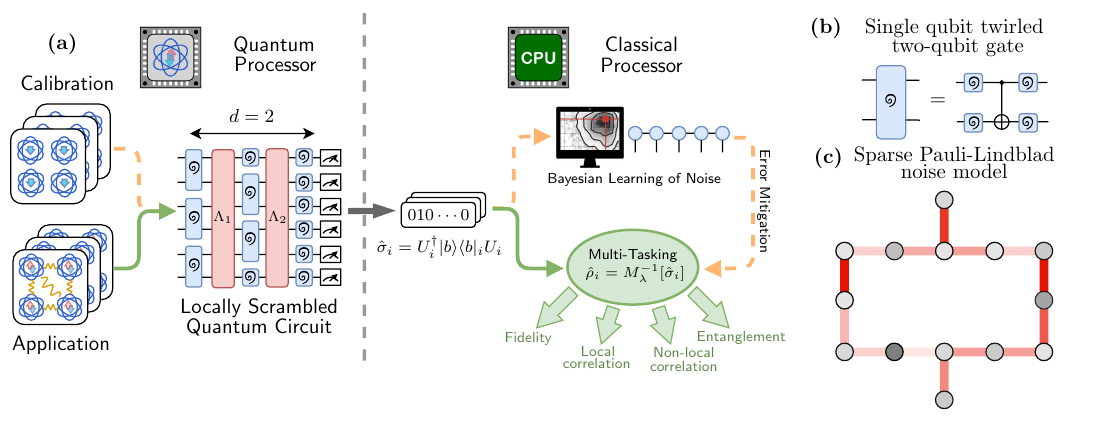}
    \caption{A schematic overview of the robust shallow shadow protocol. In (a), we show an example of our randomized measurement scheme for a shallow circuit with $d=2$, which is a brickwork circuit comprised of twirled two-qubit gates. As shown in (b), these twirled gates are CNOT gates sandwiched by single-qubit random Cliffords. Our noise model is the sparse Pauli–Lindblad model~\cite{berg2022probabilistic}, which captures realistic noise effects such as qubit cross-talk. Upon twirling via single-qubit random Clifford gates, the effective noise channel simplifies from a full Pauli-Lindblad map (which has 9 two-body terms on each edge and 3 one-body terms for each node) to the one illustrated in (c), which has only one parameters for each edge and one parameter for each node. The left half of (a) shows the dataset collection process for both calibration and application states, and the right half shows our data postprocessing method. We use a Bayesian inference algorithm to estimate the noise parameters $\lambda$ of the quantum device, and use this to error mitigate our estimates of many different observables, ranging from fidelity to entanglement entropy.}
    \label{fig:overview}
\end{figure}

We apply these theoretical insights to conduct classical shadow experiments \cite{doi:10.1126/science.abn7293,CrossPlatform,Struchalin2020E2008.05234, Zhang2021E2106.10190,2023arXiv230716882V} beyond Pauli measurements using 18 qubits on a 127-qubit superconducting quantum processor. We investigate three randomized measurement schemes using random brickwork circuits with $d \in \qty{0,2,4}$ layers of twirled CNOT gates (see \cref{fig:overview}(a)). The case $d=0$ serves as a benchmark, representing conventional noise-robust randomized Pauli measurements, while the other two schemes employ shallow random circuits of increasing depth of entangling gates. We test these schemes on application states such as the cluster state and the Affleck-Kennedy-Lieb-Tasaki (AKLT) resource state. Our analysis reveals two key findings. First, our robust protocol consistently produces accurate predictions for various physical observables across different circuit depths, demonstrating the success of our noise-resilient classical shadow approach. Second, we quantify the experimental sample complexity of the robust shallow shadows protocol. Comparing error-mitigated shallow random circuits to error-mitigated random Pauli measurements (both using our noise-robust framework), we find that shallow shadows reduce the sample complexity by up to five times for observables like fidelity and nonlocal Paulis. These improvements not only align with theoretical predictions but also persist in the presence of noise. Together, these results validate our theoretical framework and highlight the practical advantages of our protocol in enhancing the efficiency and robustness of quantum state learning.
\section{Preliminaries}

We begin by reviewing the framework of randomized measurements and classical shadows. Each randomized measurement scheme is defined by an ensemble of unitary operators $\mathcal{E}$. A random unitary $U_i$ is then sampled uniformly at random from $\mathcal{E}$ and applied to the state $\rho$, the evolved state is measured in the computational basis, and the measurement outcome $\ket{b}_i$ is recorded. This evolve-and-measure scheme is repeated $K$ times, choosing a new random unitary $U_i$ each time. This forms the randomized measurement dataset $\mathcal{D}=\qty{U_i,\ket{b}_i}_{i=1}^K$. The aim is then to predict a large number of properties of $\rho$, all using the same dataset $\mathcal{D}$, a goal that we call multitasking.

To predict these properties, the dataset must first be classically postprocessed. The first step of this is to calculate the back evolution of collapsed state $\ket{b}_i$ by $U_i$ to construct the associated \emph{classical snapshot} $\hat{\sigma}_i=U^{\dagger}_i\ket{b}_i\bra{b}_iU_i$. This is possible when there are efficient classical algorithms for simulating $U^\dagger_i$; for instance, there are well-known algorithms for calculating $\hat{\sigma}_i$ when it is a Clifford circuit~\cite{aaronson2004improved}, matchgate circuit~\cite{jozsa2008matchgates,PhysRevA.93.062332}, or finite-time Hamiltonian evolution~\cite{vidal2004efficient,PhysRevLett.93.076401}. The full set of classical snapshots can be viewed as a set of classical `shadows' of the underlying quantum state $\rho$; although a single snapshot is not enough to fully specify the state, the full set of snapshots together proves sufficient, as we show below. The linearity of quantum mechanics implies that the expectation (over both random choices of the unitary $U$ and random measurement outcomes $\ket{b}$) of the classical snapshot is related to the original state $\rho$ through a linear map $\mathcal{M}$: $\mathbb{E}[\hat{\sigma}]=\mathcal{M}[\rho]$. The precise details of this map are determined by the unitary ensemble $\mathcal{E}$. Importantly, when $\mathcal{E}$ forms a tomographically complete ensemble, $\mathcal{M}$ is an invertible map, so we can write $\rho = \mathbb{E}[\mathcal{M}^{-1}(\hat{\sigma})]$, and of course any observable of $\rho$ obeys $\Tr(\rho O)=\mathbb{E}[\Tr(\mathcal{M}^{-1}(\hat{\sigma}) O)]$. This forms the basis of randomized measurement schemes: we can estimate $\mathbb{E}[\Tr(\mathcal{M}^{-1}(\hat{\sigma}) O)]$ with an empirical average over our dataset $\mathcal{D}$, hence allowing us to estimate $\Tr(\rho O)$. Notably, the dataset $\mathcal{D}$ was not tailored to a particular observable $O$, hence we can repeat the same classical postprocessing procedure to predict a large \emph{set} of $L$ observables $\qty{O_l \mid l=1,\ldots,L}$ simultaneously. This flexibility extends beyond linear observables: since $\rho=\mathbb{E}[\mathcal{M}^{-1}[\hat{\sigma}]]$ it follows that $\tilde{\rho}^{(2)} \equiv \frac{1}{K(K-1)}\sum_{i \neq j} \mathcal{M}^{-1}[\hat{\sigma}_i] \otimes \mathcal{M}^{-1}[\hat{\sigma}_j]$ is an unbiased estimator for $\rho^{\otimes 2}$. Remarkably, this means that the dataset $\mathcal{D}$, constructed using only single-copy measurements of $\rho$, can be used to learn nonlinear properties of $\rho$: for instance, the purity can be estimated using the observable $O=\swap$ and evaluating $\Tr(O \tilde{\rho}^{(2)})$. However, the most useful property of classical shadows is that the sample complexity of achieving this multitasking has been shown~\cite{HuangKeungPreskill} to be $\mathcal{O}(\log L \cdot \max_i \norm{O_i}_{\text{sh}}^2/\epsilon^2)$, which scales \emph{logarithmically} in $L$ instead of linearly. A critical component of this scaling is the shadow norm $\norm{O_i}_{\text{sh}}^2$ of the operator $O_i$, which depends on the details of the unitary ensemble. 

The shadow map $\mathcal{M}$ and its inverse can be efficiently calculated for a large family of unitary ensembles called \emph{locally-scrambled unitary ensembles}, where the unitary ensemble satisfies the local-basis invariance condition \cite{Hu2022H2102.10132} (see \cref{app:local-scramble}). Locally-scrambled unitary ensembles are easily realized experimentally: for example, as shown in \cref{fig:overview}(b), any two-qubit gate sandwiched by random single-qubit Clifford gates satisfies the local-basis invariance condition. This procedure is also called single-qubit twirling~\cite{bennett1996mixed,knill2008randomized,magesan2011scalable}. In the following, we will call these sandwiched two-qubit gates \emph{twirled gates} for short. If the randomized quantum circuit is composed of twirled gates, as shown in \cref{fig:overview}(a), then Ref.~\cite{Bu2024C2202.03272} shows that $\mathcal{M}$ is diagonal in the Pauli basis. This means that $\mathcal{M}[P] = \omega(P) P$ for any Pauli $P$, where $\omega(P) \equiv \mathbb{E}_{U \sim \mathcal{E}}[\expval{UPU^\dagger}{0}^2]$ is called the \emph{Pauli weight}. Here and throughout, we use $\ket{0}$ as shorthand for the $n$-qubit product state $\ket{0}^{\otimes n}$. This implies that
\eqs{
&\mathbb{E}[\hat{\sigma}]=\mathcal{M}[\rho]=\dfrac{1}{2^N}\sum_{P \in \mathbb{P}_N}\omega(P)\Tr(\rho P) P \,; \\
        &\rho = \mathcal{M}^{-1}[\mathbb{E}[\hat{\sigma}]]=\dfrac{1}{2^N} \sum_{P \in \mathbb{P}_N} \omega^{-1}(P) \Tr(\mathbb{E}[\hat{\sigma}] P) P
}
where $\mathbb{P}_N$ is the $N$-qubit Pauli group. We note that the locally basis invariance of the unitary ensemble $\mathcal{E}$ effectively erases any information about the local basis for $P$. This means that~$\omega(P)$ does not depend on the exact characters in the Pauli string $P$: if the position of all non-identity operators is the same for two Pauli operators, they share the same value of Pauli weights. Therefore, despite the fact that there are $4^N$ different Pauli operators, there are only $2^N$ distinct Pauli weights. As we show later, this fact makes it natural to use a `particle-hole' basis for locally-scrambled unitary ensembles. Although there are still exponentially many Pauli weights, in the next section, we will show they can all be efficiently represented (i.e., using polynomial classical resources) with a simple tensor network representation.

So far, we have assumed that the state $\rho$ can be evolved by $U$ perfectly. However, real quantum circuits have noise, in which case the actual evolution will differ from the ideal unitary $U$. The actual evolution can be described by a channel $\sC_{U,\lambda}[\rho]$, where $\lambda$ parameterizes the noise in the evolution. For unitary ensembles in which the set of unitaries $U$ forms a group, such as those formed by the Clifford group or tensor products of single-qubit Clifford gates (i.e., random Pauli measurements), the noisy measurement channel takes a simple form when expressed in the Pauli basis~\cite{Chen2021R2011.09636,Koh2022classicalshadows,2023arXiv231019947B}:
\eqs{
&\mathcal{M}_{\lambda,\text{Clifford}} [P]=f P+\dfrac{(1-f)}{2^N}\Tr(P)I \\
    &\mathcal{M}_{\lambda,\text{Pauli}} [P]=\bigotimes_{i=1}^{N} \qty(f_i P_{i}+\dfrac{(1-f_i)}{2}\Tr(P_i)I),\label{eq:robustpauli}
}
where $P=\bigotimes_{i=1}^N P_i$ is a Pauli string, the tensor product of Pauli operators on each qubit, and $f$, $f_i$ are noise-dependent parameters. These expressions allow for efficient noise characterization and mitigation~\cite{Chen2021R2011.09636,AndrewZhao}.

However, for more general unitary ensembles, such as those involving shallow quantum circuits (which do not form a group), the form of the noisy measurement channel is not immediately apparent. For shallow quantum circuits which satisfy the local-basis invariance condition, we show the measurement channel is still diagonal in the Pauli basis: $\mathcal{M}_{\lambda}[P]=\omega_\lambda(P) P$. The noisy Pauli weights are
\begin{equation}
    \omega_\lambda(P) = \underset{U \sim \mathcal{E}}{\dsE}\qty[\expval{UPU^\dagger}{0} \expval{\sC_{U,\lambda}[P]}{0}] \label{eq:pauli-weight}.
\end{equation}
For a derivation of this, see \cref{app:local-scramble}. Compared with \cref{eq:robustpauli}, we see noisy Pauli weights are generalizations of noise parameters $f_i$ and $f$ when using noisy shallow circuits.

\begin{figure}[htbp]
    \includegraphics[width=\textwidth]{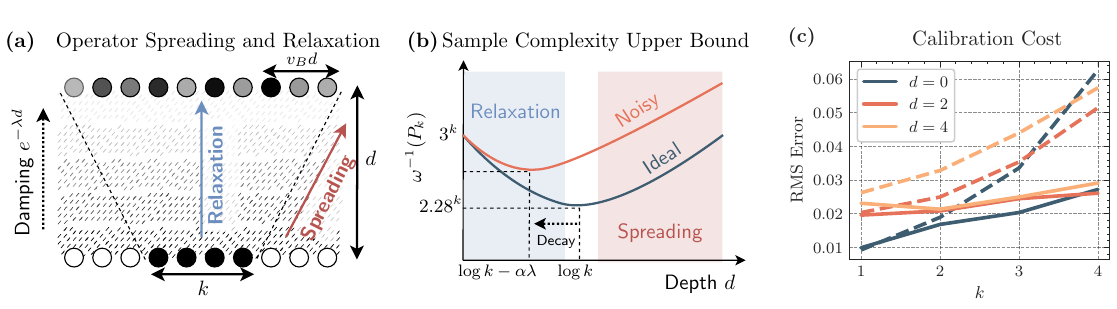}
    
    \caption{Sample complexity in robust shallow shadows (RSS) for Pauli observables with contiguous support of size $k$. In (a), we show a conceptual illustration of the three physical phenomena influencing sample complexity in the context of a classical random walk model: 1) operator spreading with a `butterfly velocity' $v_B$; 2) particle density relaxation; and 3) noise-induced damping with rate $e^{-\lambda}$. The hatched region represents the space-time domain where operator spreading and relaxation occur: as time progresses (vertically), the initial operator of size $k$ (black dots) spreads ballistically (diagonal boundaries) while simultaneously relaxing to equilibrium density (vertical blue line). These phenomena collectively determine the sample complexity required for accurate observable estimation. In (b), we illustrate the qualitative impact of noise on the sample complexity upper bound, illustrating that increased noise levels lead to a slight increase in the sample complexity upper bound and a reduction in the optimal circuit depth. This reduction is approximately linear in the noise strength $\lambda$, with a proportionality coefficient $\alpha$. This is an example of the trade-offs involved in designing a noise-robust protocol. In (c), we illustrate the RMS error of reconstructed Paulis inferred using $10^4$ different circuits with $100$ shots each, as a function of the support size $k$. The dashed lines show the RMS error associated with direct calibration, while the solid lines correspond to the Bayesian learning protocol. The results demonstrate that Bayesian learning achieves comparable or better accuracy than direct calibration across all circuit depths $d$, while requiring significantly fewer calibration shots. This improved efficiency is particularly evident for larger support sizes, where the Bayesian approach maintains stable error rates even as $k$ increases.}
    \label{fig:optimal-depth}
\end{figure}

\section{Robust shallow shadow}
\subsection{Bias-variance tradeoff}
Understanding the bias-variance tradeoff is crucial for optimizing our robust shallow shadows protocol in practice. As we increase the depth of our shallow circuits, we gain the ability to estimate a broader class of observables efficiently. However, this comes at the cost of accumulating more noise, potentially biasing our results. This tradeoff directly impacts the choice of optimal circuit depth and the overall performance of our protocol.

Since the shadow map is diagonal in the Pauli basis, the expectation of any Pauli $P$ is simply
\begin{equation}
    \Tr(\rho P) = \frac{1}{\omega_\lambda(P)} \mathbb{E}[\Tr(\hat{\sigma} P)].\label{eq:expec}
\end{equation}
Since $\abs{\Tr(\hat{\sigma} P)} \leq 1$, the inverse Pauli weight upper bounds the shadow norm of any Pauli $P$ (i.e., the sample complexity of estimating $P$): $\norm{P}^2_{\text{sh}} \leq \omega_{\lambda}(P)^{-1}$. Therefore, to upper bound the sample complexity, it suffices to lower bound the Pauli weight $\omega_{\lambda}(P)$. In the noiseless case, the sample complexity scaling of shallow shadows was numerically investigated in Refs.~\cite{Akhtar2023S2209.02093, Bertoni2022S2209.12924}. A  theoretical understanding was given by Ref.~\cite{Ippoliti_2023}, where the expectation in \cref{eq:pauli-weight} was calculated by mapping the evolution of $P$ through the circuit to a \emph{classical random walk}. The shadow norm depends on this random walk's final set of configurations. We extend this analysis \cite{Ippoliti_2023} to the noisy case. 
Our analysis assumes that the noise is single-qubit depolarizing noise with site-independent strength $\lambda$. For a detailed analysis of more general noise models, including multi-qubit correlations and non-Markovian effects, see \cref{app:noise}.

As we show, the effect of noise on this random walk is simple. While we focus our analysis on Pauli operators comprised of $k$ contiguous non-trivial characters (illustrated in \cref{fig:optimal-depth}(a)), the qualitative behaviors---operator spreading, density relaxation, and noise-induced damping---persist for non-contiguous operators. The main difference is that non-contiguous operators experience multiple independent spreading fronts, one from each cluster of non-trivial characters, leading to faster operator growth but maintaining similar asymptotic scaling. As depicted, there are three physical processes, and only the third of these depends on the noise. First, the average density of particles in the bulk will relax, owing to the fact that twirled two-qubit gates tend to reduce the weight of high weight Pauli operators (for instance, a Haar random two-qubit gate maps a weight 2 Pauli operator to a weight 1 Pauli operator with probability $\frac{3}{5}$). Second, the domain of the observable will spread (shown in red) with some butterfly velocity $v_B$ as a result of the entangling two-qubit gates which act on the edge of the Pauli's support. Finally, each particle will have a damping factor of $\exp(-\lambda)$ during the stochastic process due to the depolarizing noise. The effects of each of these three processes can be bounded, and combining all of these bounds allows us to obtain an upper bound of the shadow norm for contiguous Pauli operators, hence bounding the requisite sample complexity to estimate these operators. The details of the theoretical analysis, as well as a numerical characterization of phenomenological parameters such as $\gamma$ and $v_B$ can be found in \cref{app:optimal-depth}.

\begin{theorem}[Sample complexity and optimal circuit depth, informal]\label{thm:optimal}
    Assuming single-qubit depolarizing noise with strength $\lambda$, for $k\gg d \gg 1$, the shadow norm of a Pauli operator support over $k$ contiguous sites is upper bounded by
    \begin{equation}
        \log_3 \norm{P_k(d)}^2_{\mathrm{sh}} \leq \qty(k+d)\qty(\frac{3}{4} + \frac{\exp(-\gamma d)}{t^{3/2}} + \frac{d\lambda}{\log 3}),\label{eq:shadow-bound}
        \end{equation}
    where $d$ is the circuit depth and $\exp(-\gamma)\equiv (\frac{4}{5})^2$. The circuit depth $d^*$ that minimizes this upper bound is
    \begin{equation}
        d^*=\frac{1}{\gamma}\qty(\log \frac{k}{3 \log 3 + 4 k \lambda} + \ldots)\underset{(k\lambda\ll 1)}{=}\frac{1}{\gamma}\qty(\log \frac{k}{3 \log 3}-\frac{4k\lambda}{3\log 3} + \ldots)\label{eq:top}
    \end{equation}
    where $\ldots$ denotes subleading terms.
\end{theorem}
\cref{eq:top} shows that the theoretical optimal circuit depth is shallower for noisy circuits compared to noiseless ones, a fact which we illustrate in \cref{fig:optimal-depth}(b). Simultaneously, \cref{eq:shadow-bound} shows that the noise also increases the associated sample complexity for estimating $P_k$. However, the sample complexity bounds in this theorem are overly pessimistic, and we can do significantly better using a phenomenological model that we detail in \cref{app:optimal-depth}.

The bias-variance tradeoff highlights the need for efficient and accurate calibration methods. To address this, we develop a Bayesian inference approach that allows us to characterize and mitigate the noise in our system, striking a balance between the benefits of increased circuit depth and the challenges posed by accumulated noise.

\subsection{Efficient calibration with Bayesian inference}\label{sec:noise}

\cref{eq:expec} implies that an accurate estimate of $\omega_\lambda(P)$ is required to produce accurate estimates of Pauli expectation values. In this section, we describe an efficient Bayesian inference method to learn the Pauli weights. We begin with a cruder method: \emph{direct inference}. Simply by inverting \cref{eq:expec}, we see that if we set $\rho = \ketbra{0}$ and $P$ to be any Pauli operator formed of only $I$ and $Z$ operators, we get $\omega_\lambda(P) = \mathbb{E}[\Tr(\hat{\sigma} P)]$. We can estimate $\mathbb{E}[\Tr(\hat{\sigma} P)]$ by preparing $\ket{0}$, evolving under a random shallow circuit $U$, and measuring in the computational basis. \cref{theorem:calibration1} shows that this calibration process requires \emph{no assumptions} on the structure of noise in the circuit, and is asymptotically unbiased.

\begin{theorem}[\textbf{Informal}, Learning noisy Pauli weights]
    Given $\epsilon>0$,
    a number of $$R=\mathcal{O}\left(\epsilon^{-2} \underset{\abs{P_\alpha}\leq k}{\mathrm{max}}\norm{P_{\alpha}}^{4}_{\mathrm{sh}}\right)$$
    sample in the calibration process is enough for the following robust shallow shadow procedure to estimate any $k-$local observables to the following precision
    \begin{equation}
        \abs{\Tr(\rho O)-\widehat{O}_{\mathrm{est}}}<\epsilon\, 2^k \norm{O}_{\infty}
    \end{equation}
    with high success probability.\label{theorem:calibration1}
\end{theorem}

This direct calibration method allows us to learn individual Pauli weights efficiently. However, for predicting properties involving many noisy Pauli weights, knowing individual Pauli weights is insufficient. In practice, we require an efficient way to encode all $2^N$ Pauli weights. Although this is impossible in general, we can take advantage of certain structural features of our unitary ensemble: our random unitary ensemble consists of twirled two-qubit gates arranged in a brickwork structure. By leveraging this structural assumption, we prove that any time-independent device noise can be captured by a stochastic Pauli noise model.  The proof idea is similar to the idea behind \emph{randomized compilation}~\cite{winick2022concepts}, wherein single-qubit random gates can effectively change physical noise to stochastic Pauli noise. This result is summarized in \cref{thm:noise}, the proof of which is in \cref{app:noise}. 

\begin{theorem}[\textbf{Informal}, Noise-robust shallow shadow on time-independent noise]\label{thm:noise}
In a noise-robust shallow shadow setting, a stochastic Pauli noise model can be used in the data post-processing to effectively capture the time-independent noise that happened in the physical circuit, including coherent errors, provided that the single-qubit gate errors are small. If the noise parameters are accurately learned, the predictions given by the robust shallow shadow remain unbiased.
\end{theorem}

By integrating this simple stochastic Pauli noise model with tensor-network post-processing and Bayesian learning, we can achieve more efficient calibration and prediction, enabling the accurate prediction of a wide range of observables, including quantum fidelity. In light of \cref{thm:noise}, we model the noise channel $\Lambda$ of an $N$-qubit system from the local interactions generated by a Lindbladian $\mathcal{L}(\rho)=\sum_{k\in \mathcal{K}}\lambda_k \left(P_k\rho P_k-\rho\right)$, where $\mathcal{K}$ is a set of Pauli operators that is defined by the local interactions between qubits. Given that the interactions are assumed to be geometrically local, the size of $\mathcal{K}$ --- and thus the number of noise parameters --- scales linearly with $N$. While non-local Pauli errors may arise beyond nearest neighbors, this model has been shown to accurately describe realistic hardware noise, as demonstrated in Ref.~\cite{berg2022probabilistic}. Furthermore, in \Cref{app:noise}, we demonstrate that our protocol remains robust even in the presence of non-local correlated errors. Additionally, we extend \Cref{thm:noise} to other noise models, including non-Markovian noise, as detailed in \Cref{app:noise}. Specifically, if $U$ represents a single layer of CNOT gates, the state after this layer evolves as $\rho' = e^{\mathcal{L}}[U \rho U^\dagger]$. Since $\mathcal{L}$ is a sparse Pauli-Lindbladian, $e^{\mathcal{L}}$ enjoys a particularly simple form when acting on Pauli operators:
\[
e^{\mathcal{L}}[P] = \exp(-\sum_{k; \comm{P_k}{P} \neq 0} \lambda_k) P,
\]
indicating that $P$ acquires a damping factor based on its non-commuting Lindbladian generators~\cite{berg2022probabilistic}. We assume different noise parameters $\lambda$ for the even and odd layers of CNOT gates, and model readout errors by absorbing them into the last layer of Pauli-Lindbladian noise. We note that although the Pauli-Lindbladian model has 9 two-body terms for each neighboring pair of qubits and 3 on-site terms for each qubit, the single-qubit twirling in our shallow circuit ensemble simplifies this noise significantly. The effect of these random single-qubit gates is to twirl the noise such that for each edge, it can be parameterized by three numbers: $\lambda_{PI}, \lambda_{IP}, \lambda_{PP}$. These represent the local action of the noise channel on a Pauli operator which has support on the first qubit, second qubit, or both qubits, respectively. The first two terms can be interpreted as single-qubit depolarizing noise. 

In the post-processing, we aim to infer the `effective' values of the noise parameters $\lambda$. We emphasize that the learned noise parameters are phenomenological, in the sense that we are concerned with their values only insofar as they affect the Pauli weights $\omega_\lambda(P)$. Indeed, as we show in \cref{app:bayes}, this calibration method works well even if the device noise is not well-described by the Pauli-Lindblad model. The reason for this is that the noisy Pauli weights $\omega_\lambda(P)$ do not depend strongly on the particular structure of the device noise, hence our sparse Pauli-Lindblad ansatz is sufficiently flexible to represent the noisy Pauli weights of a wide variety of noise models.

To estimate the noise parameters $\lambda$ more efficiently, we apply the framework of Bayesian inference. Although the learning framework in probabilistic error cancellation (PEC)~\cite{berg2022probabilistic} can be used for this estimation as well, since the noise is twirled by random single-qubit Clifford gates, this results in a simplification of the effective noise channel. This simplification means that it is sufficient to use historical noise data (learned as part of PEC) as a loose prior, and to use Bayesian learning to `fine-tune' this prior. This Bayesian learning method requires less device time than the direct calibration method and Pauli-Lindblad learning, as the calibration dataset $\scD_c$ we use for the parameter estimation is relatively simple. This calibration dataset is obtained by preparing the $\ket{0}^{\otimes N}$ state and applying our shallow shadows protocol. Using this data, we construct a likelihood function $p(\scD_c | \lambda)$ based on empirical estimates of Pauli weights. We then use a log-normal prior distribution $p(\lambda)$ centered around historical noise data. The posterior distribution $p(\lambda | \scD_c)$ is sampled using Hamiltonian Monte Carlo, allowing us to infer the effective noise parameters $\lambda$. As shown in \cref{fig:optimal-depth}(c), this Bayesian approach maintains significantly lower RMS errors compared to direct calibration, particularly for Pauli operators with larger support size $k$. While the direct calibration error grows rapidly with $k$, our Bayesian method shows remarkable stability, maintaining consistent error rates across different circuit depths and operator sizes. For details of this inference method, see \cref{app:bayes}.

To further enhance our protocol's efficiency, we complement our Bayesian inference method with a tensor network-based representation of the noisy Pauli weights. This approach allows us to efficiently encode all $2^N$ Pauli weights using polynomial classical resources. By recasting the Pauli weight calculation as an expectation over a random walk in the space of Pauli operators, we construct a matrix product operator (MPO) representation of the expectation $\mathbb{E}[\mathcal{C}_2[\cdot]]$. This MPO, when applied to the all-plus state, yields a matrix product state (MPS) representation of the Pauli weights $\omega_\lambda(P)$ with bond dimension exponential in the circuit depth. For our log-depth circuits, this results in a polynomial-resource classical representation, enabling rapid estimation of a wide range of observables (see \cref{app:tensor-network} for details).

The full robust shallow shadow protocol with efficient Bayesian inference calibration is summarized in \Cref{alg1}.


\begin{alg}[label=alg1]{Noise-Robust Shallow Shadows Protocol}
    Fix a locally scrambled unitary ensemble $\mathcal{E}$. The circuits in this ensemble must be at most logarithmic depth. \vspace{-1.5mm}

\textbf{Calibration}
  \begin{enumerate}[itemsep=0mm]
      \item Prepare the $\ket{0}$ state with high fidelity and collect a calibration dataset $\scD_{c}$ by sampling from the unitary ensemble $\mathcal{E}$.
      \item Use Bayesian inference (see \cref{app:bayes}) to find an MAP estimate of true noise parameters $\lambda$.
      \item Construct a representation of the noisy Pauli weights $\omega_\lambda$ using the MPS formalism with $\lambda$, and variationally find an MPS representation of the inverse Pauli weights $\omega_\lambda^{-1}$ with gradient descent \cite{Akhtar2023S2209.02093,Bertoni2022S2209.12924}. Since empirically derived noise parameters are accounted for in $\omega_\lambda^{-1}$, these inverse Pauli weights mitigate the effects of experimental noise (e.g., qubit crosstalk, measurement error, etc.).
  \end{enumerate}
  \tcblower
  \textbf{Application}
  \begin{enumerate}[itemsep=0mm]
      \item Prepare the application state $\rho$ and collect the dataset $\scD$ by sampling from the unitary ensemble $\mathcal{E}$.
      \item Find an MPS representation (in the Pauli basis) for the observables of interest. Examples of observables that admit an efficient MPS representation include any Pauli operator, subsystem purities, and overlaps with respect to matrix product states.
      \item Infer the values of the observables of interest by contracting the appropriate MPS (see \cref{app:tensor-network}).
  \end{enumerate}
\end{alg}

\section{Experiments}
In this section, we demonstrate the effectiveness of our robust shallow shadows protocol through a series of experiments on a superconducting quantum processor. Our goals are twofold: first, to show that our protocol can accurately recover various quantum state properties in the presence of noise, and second, to quantify the sample complexity advantages of our method compared to traditional approaches. We investigate three different circuit depths and apply our protocol to multiple quantum states, including the plus state, cluster state, and AKLT resource state.

There are a number of choices that an experimentalist is free to make in \Cref{alg1}. The first is the choice of the unitary ensemble $\mathcal{E}$. As shown in the original classical shadow framework \cite{HuangKeungPreskill}, the details of this ensemble can have a drastic impact on sample complexity, depending on the observables of interest. For instance, ensembles comprised of single-qubit random Cliffords are well suited for predicting local observables, while ensembles comprised of global random Cliffords are useful for low-rank observables such as fidelity. In this section, we will provide experimental evidence showing that interpolating between these two regimes allows us to achieve, in a sense, the best of both worlds. By using shallow brickwork random Clifford circuits, we show that an extremely broad class of observables can be measured with relatively low sample complexity, including both local observables and low-rank observables such as fidelity. Concretely, we test three different ensembles which consist of $d \in \qty{0,2,4}$ layers of single-qubit twirled CNOT gates. The $d=0$ is simply the random single-qubit Clifford case initially proposed in Ref. \cite{HuangKeungPreskill}, also known simply as randomized Pauli measurements. On the other hand, as shown in \cref{fig:overview}(a), when $d=2$, our circuits contain \emph{two} layers of twirled CNOT gates (one even layer and one odd layer), so $d=4$ contains four layers of twirled CNOT gates. The other remaining degrees of freedom in \Cref{alg1} are the size of the dataset $\mathcal{D}$, the application state $\rho$, and the observables of interest. For both the calibration and application dataset, we applied $10000$ different random unitaries from $\mathcal{E}$ and took $100$ shots for each unitary circuit, keeping the random twirling gates fixed across all $100$ shots for a given unitary (see \cref{app:experiment} for experimental details). We then applied our multitasking protocol on both the plus state $\ket{+}^{\otimes 18}$ and a cluster state. The cluster state we choose is $\ket{\phi}=\prod_{i=1}^{N-1}\text{CZ}_{i,i+1}\ket{+}^{\otimes N}$, which is a ground state of a symmetry-protected topological (SPT) Hamiltonian $H=-\sum_{i}Z_{i-1}X_{i}Z_{i+1}$. Even though both are stabilizer states, we emphasize that our method doesn't rely on any special properties of the stabilizer states (later in this section, we will demonstrate this by applying our protocol to the AKLT resource state). We show below that, using the \emph{same randomized measurement dataset}, we can accurately predict a number of observables for each of these states, including fidelity, local and nonlocal Pauli observables, and subsystem purity. 
\begin{figure}
    \centering
    \includegraphics[width=0.85\linewidth]{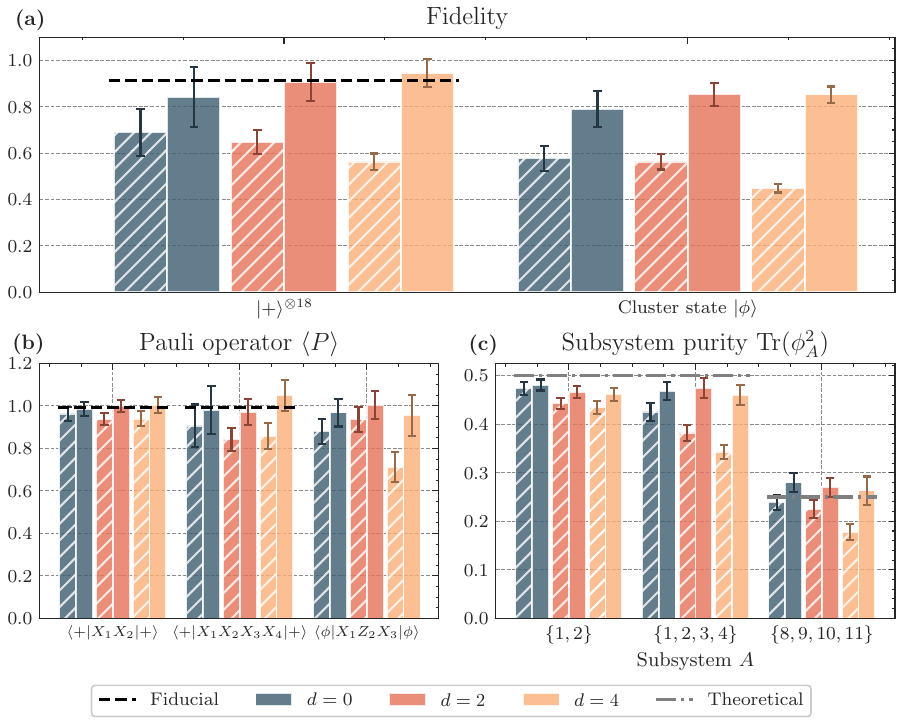}
    \caption{We apply RSS to predict multiple quantities including fidelity, Pauli observables, and subsystem purities. The hatched bars indicate recovered values without error mitigation, while the solid bars use error mitigation. In the top panel, we infer the fidelity of the experimentally prepared plus state $\ket{+}^{\otimes 18}$ and cluster state $\ket{\phi}$ with respect to the ideal (i.e., perfectly prepared) state. For the plus state, predictions agree with fiducial values obtained via direct fidelity estimation (i.e., measurement in the $X$ basis), showcasing the effectiveness of RSS in error mitigation. In contrast, predictions without error mitigation exhibit a decline in fidelity as circuit depth increases, underscoring the impact of noise. The lower left panel displays the predicted expectation values of Pauli observables, where RSS predictions maintain consistency and, for the plus state, agree with fiducial values. The lower right panel shows different subsystem purity predictions for a cluster state, illustrating how purity values are contingent upon the number of cuts within a subsystem. For instance, a subsystem in the bulk (formed by two cuts) has theoretical purity 0.25, whereas a boundary subsystem, with one cut, has a theoretical purity 0.5.}
    \label{fig:recovered}
\end{figure}

\cref{fig:recovered} summarizes our experimental results for the plus state and cluster state. The top panel shows the inferred overlap between the experimentally prepared state and the ideal application state. The hatched bars indicate recovered values without error mitigation, while the solid bars show inferred values when accounting for noise. To verify the correctness of our protocol, we also estimated some of these observables with direct measurement. For instance, the overlap of the experimentally prepared state with $\ket{+}^{\otimes 18}$ was calculated by repeatedly measuring in the $X$ basis, and the three Pauli observables were inferred similarly. The directly measured values are still subject to readout noise. Therefore, we use readout error mitigation~\cite{berg2022model} and use the mitigated results as the fiducial values (dashed black lines in \cref{fig:recovered}(a)). Both our error-mitigated shadow predictions and these fiducial measurements agree within their respective statistical uncertainties, suggesting neither method has significant unmitigated systematic bias. We note that some error-mitigated estimates (like $X_1X_2X_3X_4$ at $d=4$) slightly exceed the physical bound of 1. This is a known phenomenon in error mitigation: while our protocol guarantees unbiased estimates, statistical fluctuations combined with error mitigation can occasionally yield unphysical values. One could enforce physicality through additional constraints or renormalization, but this would introduce bias in the estimator. We choose to report the unbiased estimates directly, as their proximity to physical bounds suggests our error mitigation is working as intended without over-correction.

We can also use the dataset to predict nonlinear properties, such as subsystem purity $\Tr \psi_A^2$, where $\psi_A$ is the reduced density matrix of $\psi$ on subsystem $A$. This purity is important because it encodes information about the entanglement entropy of a state -- more specifically, it is related to the entanglement 2-R\'enyi entropy via $S_2 = -\log_2 \Tr(\psi_A^2)$. The entanglement entropy of a (perfectly prepared) cluster state is well-known. The degenerate boundary edge modes are broken in the state preparation for the cluster state, but if we cut the system into two pieces, this cut will create a degenerate edge mode, which in turn creates 1 bit of information. That is, we expect that each cut will reduce purity by a factor $\frac{1}{2}$. To be more concrete, if a subsystem is created with a single cut (e.g., subsystems $\qty{1,2}$ and $\qty{1,2,3,4}$), then the theoretically expected subsystem purity is $\frac{1}{2}$. If a subsystem is created with two cuts (e.g., subsystem $\qty{8,9,10,11}$), we expect the subsystem purity to be $\frac{1}{4}$. In \cref{fig:recovered}, we observe that indeed the predicted purity after error mitigation is close to these theoretical values. The small deviations away from theoretical values is due to imperfect state preparation: as shown in \cref{fig:recovered}, the experimentally prepared cluster state has fidelity $\sim 80\%$. The performance improvement from shallow shadows varies across observables. For some local observables like the $\{1,2\}$ subsystem purity, unmitigated $d=0$ performs as well as error-mitigated shallow shadows, indicating that shallow circuits provide minimal advantage for such local quantities. This aligns with theoretical expectations: random Pauli measurements ($d=0$) are already optimal for local observables, while shallow shadows improve sample complexity primarily for more nonlocal (e.g., the purity of subsystem $\qty{1,2,3,4}$) or low-rank observables.

\begin{figure}
    \centering
    \includegraphics[width=0.75\linewidth]{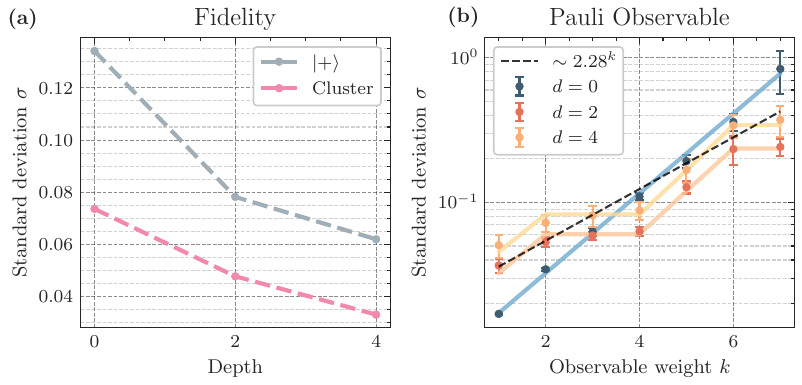}
    \caption{Since the number of samples required to achieve a certain statistical error is proportional to the variance (square of the standard deviation) of the estimator, we use the standard deviation to indicate sample complexity. (a) The standard deviation of fidelity predictions decreases with increasing circuit depth, indicating reduced sample complexity. (b) The standard deviation for estimating Pauli operator expectations is plotted as a function of the Pauli weight $k$. We observe excellent agreement between experimental data (solid dots with error bars) and theoretical predictions (solid lines). Notably, shallow circuits ($d=2,4$) exhibit favorable sample complexity scaling for higher-weight Pauli operators, outperforming the $d=0$ scaling (proportional to $3^k$), as well as the theoretical upper bound of $2.28^k$.}
    \label{fig:sample-complexity}
\end{figure}

One of the advantages of applying shallow circuits to classical shadow tomography is the reduction of sample complexity for both nonlocal and low-rank observables. To evidence this claim, we observe two competing effects. First, as shown in \cref{thm:optimal}, increased depth is well-suited for nonlocal observables. It is also well-suited for low-rank observables such as fidelity, since we converge to the global Clifford ensemble (which is optimal for low-rank observables) with increased depth. This improved suitability reduces the standard deviation of the inferred values for these observables at larger depths: this effect is particularly evident for fidelity, as highlighted in \cref{fig:sample-complexity}. 

However, this effect competes against the increasing effects of noise. As can be seen in \cref{fig:recovered}, when error mitigation is turned off, the recovered values become more biased with increasing depth $d$. This depth-dependent bias is most pronounced for fidelity because it is a global observable sensitive to all qubits, accumulating errors from the entire circuit. In contrast, local observables like $X_1X_2$ are affected only by errors in their local region, making them more robust to increasing circuit depth. As discussed in \cref{sec:noise}, this noise-induced bias can be corrected using Pauli weights $\omega_\lambda^{-1}$ that have been appropriately calibrated. However, this comes at the cost of slightly increased variance, which in turn increases the sample complexity. As with the bias, this increase in sample complexity also grows with depth $d$.

Using experimental data, we quantitatively study the competition between these two effects in \cref{fig:sample-complexity}. Using bootstrap estimates of the standard deviation $\sigma$ of our recovered expectation values, we can study the effects of increased depth on a variety of observables, including fidelity and a set of Pauli observables with increasing weight. We observe that for high-weight Pauli observables, the two competing effects find an optimal tradeoff at $d=2$. Although upper bounds from \cref{thm:optimal} predict a standard deviation $\sim 2.28^k$ at the optimal depth in an ideal setting (already a significant improvement over $\sigma \sim 3^k$, which is the scaling for $d=0$), we find that in practice, we can do much better than this, as evidenced by the $d=2$ line compared to $2.28^k$ dashed line. Our MPS formalism allows us to calculate an \emph{exact} prediction for the standard deviation of any Pauli observable, shown in solid lines, demonstrating excellent agreement with the bootstrapped standard deviations. Turning to fidelity estimation, we see, unlike Pauli observables, $d=4$ is a strict improvement over $d=2$. This contrasting behavior reflects the different scaling shown in \cref{fig:optimal-depth}(b): while Pauli observables reach optimal sample complexity at moderate depths before noise degradation dominates, fidelity estimation continues improving with depth despite increased noise. This is expected, as random global Clifford shadows is the optimal setting for fidelity estimation in the noiseless limit \cite{HuangKeungPreskill}. Regardless, for Pauli observables with contiguous support (like those shown here) and fidelity, using a shallow depth ensemble with error mitigation is always strictly better than using a $d=0$ ensemble. While our experiments focus on contiguous Pauli strings for simplicity, theoretical analysis suggests similar advantages hold for non-contiguous strings (see \cref{app:noise}), though the optimal circuit depth may vary with the Pauli support pattern. 
\begin{figure}
    \centering
    \includegraphics[width=0.95\linewidth]{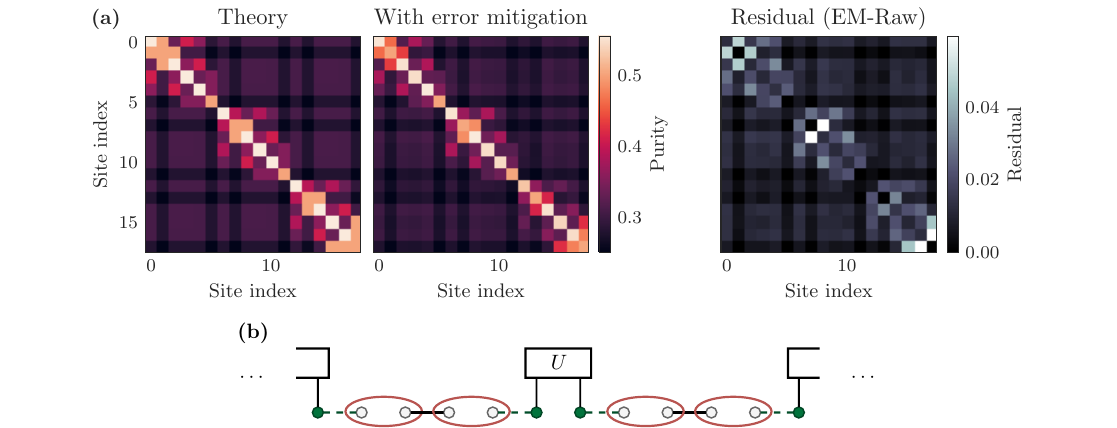}
    \caption{Prediction of subsystem purity in the AKLT resource state using RSS. In (a), we demonstrate how the RSS method can use a single dataset to concurrently predict the purity of all subsystems up to two qubits within AKLT resource states. We show theoretical predictions (left), experimental results with error mitigation (center), and the residual difference between mitigated and unmitigated results (right). The residual plot reveals that error mitigation systematically increases the predicted purities, bringing them closer to theoretical values, with the strongest corrections appearing in the three distinct AKLT clusters. The values at $(i,j)$ represent the purity of the reduced density matrix $\Tr(\rho_{ij}^2)$. The AKLT resource state has three clusters, each representing a smaller AKLT state with two spin-1 particles before fusion measurement; experimental predictions clearly show this pattern as well, and closely align with theoretical predictions. In (b), we show a schematic of the AKLT resource state before fusion measurements are applied to prepare the AKLT state.}
    \label{fig:entanglement}
\end{figure}

To further demonstrate the versatility of our protocol, we apply it to a more complex quantum state: the AKLT resource state. This state is particularly interesting as it is not a stabilizer state~\cite{PRXQuantum.4.020315}, unlike the plus and cluster states we examined earlier. By characterizing this state, we showcase our protocol's ability to handle a broader class of quantum states, including those relevant to quantum simulation and quantum computational supremacy experiments. Specifically, we applied RSS with $d=4$ to predict the purity of all size 1 and 2 subsystems for the AKLT resource state on 18 qubits. As shown in \cref{fig:entanglement}(b), this state consists of 3 small clusters of AKLT states, which can eventually be merged into a 6-qubit AKLT state. The small clusters are knit together into a single global AKLT state by applying Bell measurements on the edge qubits of adjacent small AKLT states, and applying a correction conditioned on the measurement outcome (this process is known as applying `fusion measurements'). However, in this work, we do not apply these fusion measurements, as this requires measurement feedforward, which is experimentally difficult to implement. Instead, we prepare the AKLT resource state on superconducting qubits and characterize its entanglement structure, prior to fusion measurements, using RSS. In \cref{fig:entanglement}(a), we show theoretical and experimentally recovered values for every 1- and 2-qubit subsystem purity of the resource state. The theoretical predictions assume perfect state preparation and noiseless evolution, providing an idealized benchmark against which we can compare our error-mitigated experimental results. The mean relative error (compared to the theoretically predicted values) for the recovered purities without error mitigation is $6.4\%$, while the mean relative error with error mitigation is $3.2\%$, further evidencing the efficacy of our error mitigation method. Notably, the residual difference between mitigated and unmitigated results is consistently positive across all subsystems, indicating that our error mitigation protocol systematically reduces excess entropy introduced by experimental noise, particularly within the three AKLT clusters. Comparing the recovered purities shows excellent agreement with exact theoretical values: as expected, we clearly see three distinct entangled clusters in the experimental data, representing each of the local AKLT states.


\section{Outlook}
In this work, we executed an unbiased randomized measurement experiment that went beyond random Pauli measurements for large quantum systems. In these experiments, we showed that our \emph{robust shallow shadow} protocol can efficiently recover unbiased estimates for a wide array of observables on unknown quantum states, even under the presence of noise. While our Bayesian noise characterization technique is general, we demonstrate its effectiveness using a sparse Pauli-Lindblad noise model well-suited to the IBM device's heavy-hex architecture—other quantum platforms may require different device-specific noise models. We do this by providing a general noise characterization technique based on Bayesian inference that can account for realistic noise effects, such as qubit crosstalk and measurement error. Having characterized the noise on our device, we then introduced an efficient tensor network-based postprocessing technique that is naturally able to account for the effects of this noise. We provide evidence that this error mitigation technique is effective by demonstrating that our protocol gives unbiased predictions of low-rank observables (e.g., fidelity), nonlocal Pauli observables, and even non-linear observables (e.g., subsystem purity) for a number of application states, including the cluster state and the AKLT resource state. Not only is our protocol able to recover unbiased estimators of these observables, we show that the standard deviation of these estimators improves with shallow circuits compared to $d=0$, though the optimal depth varies by observable type. While fidelity estimates continue improving through $d=4$, Pauli observables achieve minimum variance around $d=2$ before noise effects begin to dominate. That is, we show how going beyond random Pauli measurements can give rise to improved sample complexities. Furthermore, we developed a theoretical framework that not only predicted improvements in sample complexity that agreed well with the empirically observed improvements, but our framework also showed that shallow shadows remain information-theoretically optimal, even under the presence of noise. Our new theoretical insights, combined with the experimental validation of our protocol, further underscores the practical relevance and effectiveness of our approach. The success of these experiments not only validates the theoretical underpinnings of our protocol but also showcases its potential for real-world quantum computing applications. Finally, we note that the framework developed in Ref.~\cite{modelviolation} could be used to characterize out-of-model errors in our noise model. When there is a significant mismatch between the effective Pauli noise model and the actual device noise, the Bayesian calibration procedure may fail, leading to biased predictions. In such cases, the \emph{direct calibration algorithm} outlined in \cref{sec:noise} provides a model-agnostic method to estimate the relevant Pauli weights. In \cref{app:bayes}, we evaluate the performance of our local Pauli Lindbladian noise model in the presence of additional correlated non-local three-qubit noise. Across a wide range of strengths for these three-qubit errors, our framework remains robust, delivering unbiased predictions within statistical errors. Furthermore, in our experiments, comparisons between error-mitigated shadow predictions and fiducial values obtained from other direct measurement schemes show no deviations beyond statistical uncertainties.

We have presented a protocol that maintains the advantages of classical shadow tomography even in noisy quantum systems. This noise resilience is particularly important for the many applications of classical shadows in quantum machine learning, quantum chemistry, and quantum many-body physics. The randomized measurements dataset serves as a succinct classical description of a quantum state, and our robust protocol ensures these measurements remain reliable on real devices. By predicting many different Pauli observables efficiently, one can do unsupervised learning of conserved quantities, symmetry, and phases of matter for quantum many-body systems \cite{2023arXiv230900774Z,2024arXiv240102940L,doi:10.1126/science.abk3333}. A similar approach can be used for ansatz-free Hamiltonian learning and quantum device benchmarking \cite{Andi,PRXQuantum.2.010102,PhysRevLett.124.160502}. Measuring many low-rank observables simultaneously could also lead to new applications. For example, estimating the low eigenenergy spectrum is important for quantum many-body physics and quantum chemistry. Combining the idea of dynamic mode decomposition \cite{2023arXiv230601858S} and measuring many low-rank observables simultaneously, one could have a better convergence rate in predicting eigenenergies \cite{Shen24}. 

To fully realize these potential applications and validate the broad applicability of our approach, future work should focus on extending these experiments to a wider range of quantum computing platforms and larger system sizes. While our current results demonstrate the effectiveness of robust shallow shadows on superconducting qubits, exploring its performance on other architectures such as trapped ions, neutral atoms, or photonic systems would be valuable. This cross-platform validation would not only further confirm the generality of our approach but also potentially reveal platform-specific optimizations. Additionally, investigating the scalability of our method to larger system sizes would be crucial for establishing its utility in future large-scale quantum applications.

In parallel with these experimental directions, it would be interesting to use modern machine learning techniques to further improve the inference and predictions of the proposed method \cite{2023arXiv230917368L} and also improve the sample complexity by tailoring unitary ensembles to special observables of interest \cite{2024arXiv240118071F,Van-Kirk2022H2212.06084}. Lastly, by integrating the tensor-network representation of the quantum noise model \cite{learning_of_QPT,2023arXiv231208454W,PhysRevResearch.6.033217} with a robust shallow shadow dataset, it may be possible to mitigate quantum errors that occur during quantum simulation through data post-processing. This approach has the potential to improve the accuracy of quantum simulations on near-term quantum devices \cite{TEM}.

\textit{Note added} --- During the completion of this manuscript, we became aware of a related but independently developed work taking a randomized benchmarking approach to mitigate noise in randomized measurements~\cite{eisert2024noise}.

\section{Acknowledgements}
We would like to thank Gefen Baranes, Pablo Bonilla, Nazl\i \ U\u{g}ur K\"oyl\"uo\u{g}lu, and Varun Menon for insightful discussions. SFY and HYH thank the NSF for funding through the Q-IDEAS HDR Institute (OAC-2118310), the CUA PFC (PHY-2317134), and DARPA through their IMPAQT Program (HR0011-23-3-0023).
YZY is supported by a startup fund from UCSD. We acknowledge the KITP program ``Quantum Many-Body Dynamics and Noisy Intermediate-Scale Quantum Systems'' where the research collaboration was initiated.

\section{Data Availability}
Source data are available for this paper. All other data supporting the plots within this paper and other study findings are available from the corresponding author upon reasonable request.

\section{Code Availability}
The code used in this study is available from the corresponding author upon request.

\section{Author Contributions}
H.Y.H. and A.G. developed the theory of robust shallow shadows and wrote the code to analyze experimental data. H.Y.H., Y.Z.Y., S.F.Y., A.S., and Z.M. designed the application and experiments. S.M. conducted the experiments and collected the randomized measurement dataset. A.G. conducted the data postprocessing of the experimental data. H.R., Y.Z. and D.S.W. contributed to the initial discussion and setup of this work. All authors contributed substantially to writing the manuscript.

\bibliographystyle{apsrev4-2} 
\bibliography{refs}

\clearpage
\appendix
\begin{center}
	\noindent\textbf{Supplementary Material}
	\bigskip
		
	\noindent\textbf{\large{}}
\end{center}
\section{Locally scrambled classical shadow with general quantum channel\label{app:local-scramble}}

In the following discussion, we consider circuits where all two-qubit and multi-qubit gates are sandwiched between random single-qubit Clifford gates - an operation known as ``single-qubit twirling''. We call any circuit composed entirely of such twirled gates a locally scrambled circuit. Mathematically, a unitary ensemble defined by probability distribution $P(U)$ over unitaries is called \emph{locally-scrambled} if it satisfies the \emph{local-basis invariance} condition
\begin{equation}
   P(U)=P(VU)=P(UV)\qc \forall V\in U(2)^N,
\end{equation}
where $V \in U(2)^N$ denotes any tensor product of $N$ single-qubit unitaries~\cite{Hu2023C2107.04817}. This invariance means the ensemble statistics remain unchanged under left or right local basis transformations. Since our circuits are composed of gates twirled by random single-qubit Clifford gates, the total circuit ensemble inherits this local scrambling property.

In classical shadow experiments, the unitary evolution channel $U$ is sampled anew from the unitary ensemble $\scE_U$ for each measurement. We use $\sC_{U,\lambda}$ to denote the \emph{noisy measurement channel} and $\sC_{U,\mathrm{recon}}$ to denote the classical shadow \emph{reconstruction channel}, with the subscript $U$ explicitly showing the dependence on the ideal quantum circuit for each randomized measurement. While a common choice for $\sC_{U,\mathrm{recon}}[\cdot]$ is simply the unitary channel itself (i.e., $\sC_{U,\mathrm{recon}}[\cdot]=U \cdot U^{\dagger}$), knowledge of the noise channel can be incorporated to achieve improved performance. For instance, if the Kraus operators $\{K_i\}$ of the noise channel are known, one could reconstruct the classical shadows as $\hat{\sigma}_i=\sum_i K^{\dagger}_i \ket{b}\bra{b}K_i$.

Using this notation, the classical shadows are given by
\eqs{
\hat{\sigma}_{U,b}=\sC^{\dagger}_{U,\mathrm{recon}}[\ket{b}\bra{b}],
}
with corresponding probability
\eqs{
p(\hat{\sigma}_{U,b}|\rho)=\Tr(\ket{b}\bra{b}\sC_{U,\lambda}[\rho]).
}
Under the locally scrambling assumption, the measurement channel is diagonal in the Pauli basis:
\eqs{
\scM[\rho]&=\underset{U}{\dsE}\sum_b \hat{\sigma}_{U,b}p(\hat{\sigma}_{U,b}|\rho)\\
& = \dfrac{1}{2^n}\sum_{P \in \mathbb{P}_n}\omega(P)\Tr(\rho P) P,
}
where $\omega(P)$ is the Pauli weight of Pauli string $P$, given by
\begin{equation}\label{eq:a5}
    \omega(P)=\dfrac{1}{2^n}\sum_b \underset{U}{\dsE}\left(\Tr(\sC^{\dagger}_{U,\mathrm{recon}}[\ket{b}\bra{b}]P)\Tr(\sC^{\dagger}_{U,\lambda}[\ket{b}\bra{b}]P)\right).
\end{equation}

Since both quantum channels $\sC^{\dagger}_{U,\mathrm{recon}}$ and $\sC^{\dagger}_{U,\lambda}$ possess local basis invariance, the dependence on the specific bit-string state $\ket{b}$ vanishes. This allows us to simplify the Pauli weight as
\eqs{
\omega(P)&= \underset{U}{\dsE}\Tr\left(\sC^{\dagger}_{U,\mathrm{recon}}[\ket{0}\bra{0}^{\otimes N}]P\right)\Tr\left(\sC^{\dagger}_{U,\lambda}[\ket{0}\bra{0}^{\otimes N}]P\right)\\
&=\underset{U}{\dsE}\Tr\left(\ket{0}\bra{0}^{\otimes N}\sC_{U,\mathrm{recon}}[P]\right)\Tr\left(\ket{0}\bra{0}^{\otimes N}\sC_{U,\lambda}[P]\right).
}

By defining $\sC_{U}=\sC_{U,\mathrm{recon}}\otimes \sC_{U,\lambda}$, we can write the Pauli weight more compactly as
\eqs{
\omega(P)&= \underset{U}{\dsE}\Tr\left(\ket{0}\bra{0}^{\otimes 2N}\sC_U [P^{\otimes 2}]\right).
}

In practice, post-processing typically uses the ideal unitary evolution ($\sC_{U,\mathrm{recon}}[\cdot]=U\cdot U^{\dagger}$) while the physical channel in the experiment is noisy. The resulting Pauli weight
\eqs{
\omega(P)=\underset{U}{\dsE}\Tr\left(\ket{0}\bra{0}^{\otimes N}UPU^{\dagger}\right)\Tr\left(\ket{0}\bra{0}^{\otimes N}\sC_{U,P}[P]\right)
\label{eq:pauli-weight2}
}
has a close connection to randomized benchmarking \cite{choi_benchmark}. In \cref{app:tensor-network}, we develop a tensor-network based method for efficiently calculating this Pauli weight using polynomial classical resources, and demonstrate its application to estimating nonlocal properties such as quantum state fidelity.

\section{Theory of robust shallow shadows for general noisy channel}\label{app:noise}

In the previous section, we derived the measurement channel and reconstruction map in a general setting with only one requirement: the noisy quantum channel ensemble must be \emph{locally scrambled}. This assumption is straightforward to satisfy in practice by adding random single-qubit rotations before and after the quantum circuit. A key insight is that the measurement channel is diagonal in the Pauli basis:
\eqs{
\scM[\rho]=\underset{U}{\dsE}\sum_b \hat{\sigma}_{U,b}p(\hat{\sigma}_{U,b}|\rho) = \dfrac{1}{2^n}\sum_{P \in \mathbb{P}_n}\omega(P)\Tr(\rho P) P.
}

Using the Pauli-transfer matrix (PTM) or the Liouville representation, we can express the shadow map as
\begin{equation}
    \widehat{\mathcal{M}}=\sum_{P \in\mathbb{P}_N}\omega(P)\rket{P}\rbra{P}.\label{eq:ptm}
\end{equation}
Furthermore, due to the local basis invariance of the unitary ensemble, the eigenvalues $\omega(P)$ are identical for any two Pauli operators that share the same position support of non-identity matrices. For example, $\omega(\sigma_{IZIY})=\omega(\sigma_{IXIY})=\omega(\sigma_{IZIZ})$. This equivalence substantially reduces the number of unique eigenvalues from $4^N-1$ to $2^N-1$.

We will now demonstrate that \cref{eq:ptm} generalizes the traditional noise-robust shadow tomography with random Pauli and random Clifford measurements \cite{Chen2021R2011.09636}. When the ideal unitary ensemble forms a group $\mathbb{G}$, we can show that
\eqs{
\widehat{\mathcal{M}}=\sum_{\lambda\in R_{\mathbb{G}}}\dfrac{\text{Tr}\left(\widehat{\mathcal{M}}_{z}\Lambda\Pi_{\lambda}\right)}{\text{Tr}\left(\Pi_{\lambda}\right)}\Pi_{\lambda},
}
where $R_{\mathbb{G}}$ is the set of irreducible representations of group $\mathbb{G}$, $\Pi_\lambda$ is the projector to the invariant subspace, $\Lambda$ represents the noise channel, and $\widehat{\mathcal{M}}_z=\qty[\rket{\sigma_0} \rbra{\sigma_0}+\rket{\sigma_z} \rbra{\sigma_z}]^{\otimes N}$ represents measurement in the Pauli-Z basis. We will examine the standard robust shallow results for $\mathbb{G}=\text{Cl}(2^N)$ and $\mathbb{G}=\text{Cl}(2)^{\otimes N}$, and then explain why our construction offers greater generality.

\begin{figure}[htbp]
    \centering
    \includegraphics[width=1\linewidth]{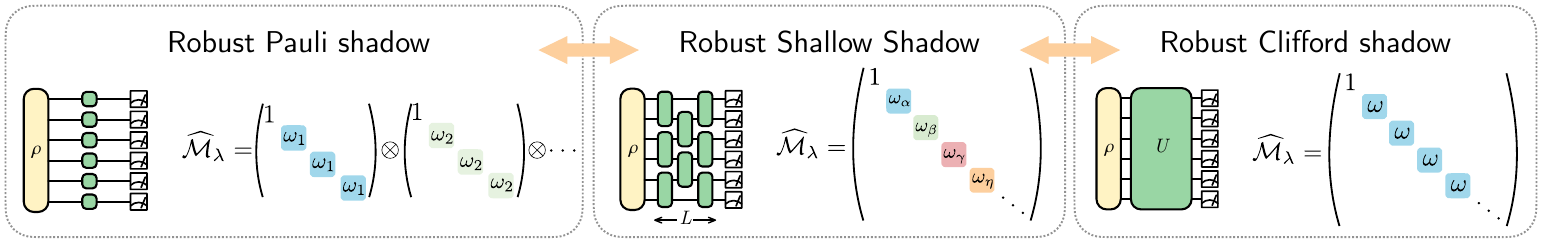}
    \caption{Noise-robust shadow map $\mathcal{M}_\lambda$ of noisy quantum circuits in the Liouville representation with Pauli basis or PTM representation. (Left) The noisy shadow map of robust Pauli measurement, where $\omega_i$ are parameters that depend on single-qubit noise. (Middle) The noisy shadow map of robust shallow shadow, where $\omega_i$ are noise-dependent parameters. Here, $\omega_i=\omega_j$ if Pauli string operators $P_i$ and $P_j$ share the same positions for non-trivial Pauli operators. (Right) The noisy shadow map of robust Clifford shadow, where there is only one noise-dependent parameter $\omega$ for all Pauli string operators, related to the averaged circuit noise in randomized benchmarking \cite{PRXQuantum.3.020357}.}
    \label{fig:liouville}
\end{figure}

It is well established that the $N$-qubit Clifford group has two irreducible representations with projectors $\rket{\sigma_0}\rbra{\sigma_0}$ and $\id-\rket{\sigma_0}\rbra{\sigma_0}$. The noisy shadow map for robust Clifford shadow can thus be expressed as 
\eqs{
\widehat{\mathcal{M}}_{\mathrm{Clifford}}=\rket{\sigma_0}\rbra{\sigma_0}+f (\id-\rket{\sigma_0}\rbra{\sigma_0}),
}
where $f\in \mathbb{R}$. In the noiseless case, $f=1/(2^N+1)$, while in the noisy case, $f$ must be learned from benchmarking experiments \cite{Chen2021R2011.09636}. In the PTM representation, this can be written as $$\widehat{\mathcal{M}}_{\mathrm{Clifford}}=\sum_{P\in \mathbb{P}_N}(1+(\omega-1)\delta_{P,\sigma_0})\rket{P}\rbra{P}$$ with $\omega = f$, as visualized in the right panel of \cref{fig:liouville}.

For random Pauli measurement, the unitary ensemble also forms a group with $2^N$ irreducible representations \cite{Chen2021R2011.09636,PhysRevLett.109.240504}, yielding the noisy shadow map
\eqs{
\widehat{\mathcal{M}}_{\mathrm{Pauli}}=\sum_{z\in\{0,1\}^N}f_z \Pi_z,
}
where $\Pi_z=\otimes_{z_i}\Pi_{z_i}$ with 
\eqs{
\Pi_{z_i} = 
\begin{cases}
    \rket{\sigma_0}\rbra{\sigma_0}, & z_i = 0, \\
    \id - \rket{\sigma_0}\rbra{\sigma_0}, & z_i = 1.
\end{cases}
}

In the noiseless case, $f_z=3^{-|z|}$, while in the noisy case, relevant $f_z$ values can be learned through direct benchmarking experiments \cite{Chen2021R2011.09636}. In the PTM representation, the noisy shadow map becomes $$\widehat{\mathcal{M}}_{\mathrm{Pauli}}=\otimes_{i=1}^{N}(1+(\omega_i - 1)\delta_{P_i,\sigma_0})\rket{P_i}\rbra{P_i}$$, where $P_i$ is the Pauli operator on the $i$-th qubit, as shown in the left panel of \cref{fig:liouville}.

The middle panel of \cref{fig:liouville} visualizes the PTM representation of the noisy shadow map for locally-scrambled classical shadow (shallow shadow). This formulation demonstrates that robust shallow shadow generalizes previous robust Pauli and Clifford shadow tomography approaches, with these methods representing the zero circuit depth and deep circuit depth limits, respectively. Importantly, our robust shallow shadow formulation does not require the randomized circuit ensemble to form a \emph{group}, making it applicable to a broader family of randomized measurement circuits, including shallow circuits and Hamiltonian dynamics twirled by single-qubit random gates.



\subsection{Calibration sample complexity}

\begin{theorem}[Learning noisy Pauli weights]
    Given $\epsilon,\delta>0$, and an integer $k\leq N$, 
    a number of $$R=\mathcal{O}\left(\epsilon^{-2} \underset{\mathrm{loc}(P_\alpha)\leq k}{\mathrm{max}}\norm{P_{\alpha}}^{4}_{\mathrm{sh}}\left(k\ln N+\ln(\delta^{-1})\right)\right)$$
    sample in the calibration process is enough for the following robust shallow shadow procedure to estimate any $k-$local observables to the following precision 
    \begin{equation}
        \abs{\Tr(\rho O)-\widehat{O}_{\mathrm{est}}}<\epsilon 2^k \norm{O}_{\infty}
    \end{equation}
    with success probability at least $1-\delta$.\label{theorem:calibration1_formal}
\end{theorem}

\begin{proof}
    Any $k$-local operator can be written as $O=\widetilde{O}\otimes I_{2^{N-k}}=\sum_{\widetilde{a}\in\mathbb{Z}_{2}^{2k}}c_{\widetilde{a}}P_{\widetilde{a}}\otimes I^{\otimes (N-k)}$.
    First, we show that the direct calibration method gives unbiased estimates of noisy Pauli weights, i.e. $\mathbb{E}[\bra{b}UP_{\alpha,z}U^{\dagger}\ket{b}]=\omega_{\lambda}(P_\alpha)$, where $P_\alpha\in \mathbb{P}_N$ is a Pauli string, and $P_{\alpha,z}$ is a Pauli string generated by $P_\alpha$ by replacing any non-trivial Pauli operator to Pauli Z operator. Since the quantum circuit ensemble is local-basis invariant, we have $\omega_\lambda(P_\alpha)=\omega_\lambda(P_{\alpha,z})$. Since $\abs{\expval{U P_{\alpha,z} U^\dagger}{b}} \leq 1$, we can we can apply a Chernoff bound to see that
    \begin{equation}\label{eq:b5}
        \Pr(\abs{\widetilde{\omega}_\lambda(P_\alpha) - \omega_\lambda(P_\alpha)} \geq \gamma) \leq \delta,
    \end{equation}
    where the sample mean estimator $\widetilde{\omega}_\lambda(P_\alpha)$ uses $n \leq 34 \gamma^{-2} \cdot \log(\delta^{-1})$ samples. Then, the prediction error can be bounded as
    \begin{equation}
        \begin{gathered}
            \abs{\Tr(\widetilde{\rho}O)-\Tr(\rho O)}=\frac{1}{2^N}\abs{\sum_{P_\alpha\in \mathbb{P}_N}(\widetilde{\omega}_{\lambda}(P_\alpha)^{-1}\omega_\lambda(P_\alpha)-1)\Tr(\rho P_\alpha)\Tr(P_\alpha O)} \\
            \leq \frac{1}{2^N}\max_{\mathrm{loc}(P_\alpha)\leq k} \abs{\widetilde{\omega}_{\lambda}(P_\alpha)^{-1}\omega_\lambda(P_\alpha)-1} \sum_{P_\alpha}\abs{\Tr(P_\alpha O)}.
        \end{gathered}
    \end{equation}
    We subsequently bound
    \begin{equation}
        \frac{1}{2^N}\sum_{P_\alpha\in \mathbb{P}^N}\abs{\Tr(P_\alpha O)}=\sum_{\widetilde{a}\in \mathbb{Z}_{2}^{2k}}\abs{c_{\widetilde{a}}}\leq \sqrt{4^k}\sqrt{\sum_{\widetilde{a}\in \mathbb{Z}_{2}^{2k}}c^2_{\widetilde{a}}}=2^k\sqrt{\frac{\Tr(\widetilde{O}^2)}{2^k}}\leq 2^k \norm{\widetilde{O}}_{\infty}=2^k \norm{O}_{\infty}
    \end{equation}
    If we set $\gamma=\epsilon|\omega_\lambda(P_\alpha)|$ in \cref{eq:b5}, then we can ensure the probability that $|\widetilde{\omega}_\lambda(P)\omega_\lambda(P)^{-1}-1|\leq \epsilon$ is at least $1-\delta$. There are at most $N^k$ $k$-local Pauli strings, and we can set $\delta\rightarrow \delta/N^k$ and apply the union bound. Then we get the final sample complexity:
    \eqs{n=64(k\ln N+\ln(2\delta^{-1}))\epsilon^{-2}\max_{\mathrm{loc}(P_\alpha)\leq k}\left|\frac{1}{\omega_\lambda(P_\alpha)}\right|^2 =64(k\ln N+\ln(2\delta^{-1})) \epsilon^{-2}\max_{\mathrm{loc}(P_\alpha)\leq k}\norm{P_\alpha}^4_{\mathrm{sh}},}
    which will ensure the estimation of any $k-$local observable to precision $\epsilon 2^k \norm{O}_{\infty}$ with success probability at least $1-\delta$.
\end{proof}

\subsection{Effective noise model}\label{app:effective}
In the following, we summarize various effective noise models, both Markovian and non-Markovian, after twirling by single-qubit Clifford gates. We assume that single-qubit gates are either ideal or subject to gate-independent errors. As demonstrated in \Cref{app:experiment}, the noise associated with single-qubit gates is orders of magnitude smaller than that of two-qubit gates or measurements, validating the practicality of this assumption.
\begin{figure}[htbp]
    \centering
    \includegraphics[width=1\linewidth]{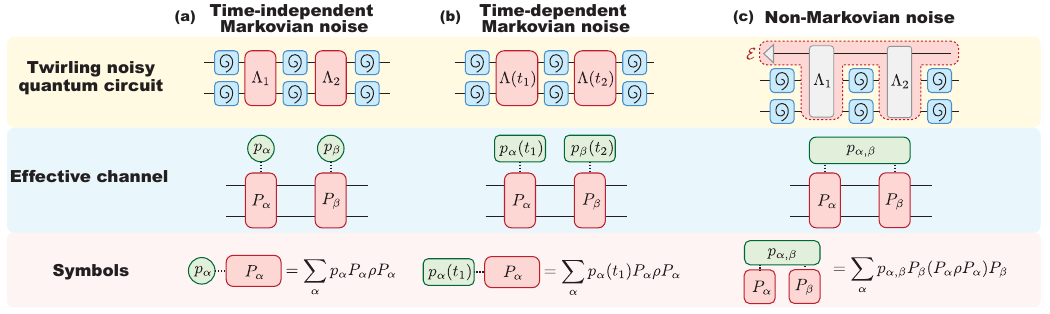}
    \caption{Effective channel after twirling with single-qubit random Clifford gates. (a) Under time-independent Markovian noise, the effective channel is a stochastic Pauli noise channel with fixed Pauli probabilities, independent across layers. (b) For time-dependent Markovian noise, the effective channel is a stochastic Pauli noise channel, but with time-dependent Pauli probabilities, while different layers remain independent. (c) Under non-Markovian noise, the effective channel exhibits stochastic Pauli noise with joint Pauli probabilities correlated across layers.}
    \label{fig:effective_channel_twirling}
\end{figure}

\begin{lemma}[Twirling of time-independent Markovian noise \cite{SymmetrizedNoise,PhysRevA.94.052325}]\label{lemma:RC1}
    Randomly sampling the twirling gates independently in each round tailors the noise at each time step (except the last) into stochastic Pauli noise with independent Pauli error probabilities when the noise on the `easy gates' (i.e., single-qubit gates) is gate-independent.
\end{lemma}

\begin{theorem}[Noise-robust shallow shadow on time-independent Markovian noise]\label{thm:nonmarkovian-noise_app}
In a noise-robust shallow shadow setting, a time-independent stochastic Pauli noise model can be used in the data post-processing to effectively capture the effects of time-independent Markovian noise that occurred in the physical circuit. If the noise parameters are accurately learned, the predictions given by the robust shallow shadow remain unbiased.
\end{theorem}

\begin{proof}
In the previous section, we showed that the noisy shadow map is fully determined by the noisy Pauli weights, expressed as
$$\omega_{\lambda}(P)= \underset{U}{\dsE}\Tr\left(\ketbra{0}^{\otimes N}U^{\dagger}P U\right)\Tr\left(\ketbra{0}^{\otimes N}\sC_{U,\lambda}[P]\right),$$
where the noisy measurement channel, $\sC_{U,\lambda}$, consists of a shallow circuit with noisy two-qubit gates, twirled by single-qubit random Clifford gates, where we assume the single-qubit gate error is small and can be neglected compared to the two-qubit gate error and readout error. 

For each measurement circuit, we can insert additional single-qubit random Clifford gates $V$ after each single-qubit gate, and let $(U, V)$ denote this new quantum circuit. This operation does not alter the value of the noisy Pauli weights, so we have
\begin{equation}
    \omega_{\lambda}(P) = \dsE_{U,V}\Tr\left(\ket{0}\bra{0}^{\otimes N}\tilde{U}^{\dagger}P \tilde{U}\right)\Tr\left(\ket{0}\bra{0}^{\otimes N}\sC_{(U,V),\lambda}[P]\right),
\end{equation}
where $\tilde{U}$ is the updated circuit after inserting the additional random single-qubit Clifford gates $V$ into the original measurement circuit $U$. Since random single-qubit Clifford gates form a 3-design for each qubit, they effectively symmetrize the noise. By Lemma.\ref{lemma:RC1}, it transforms the time-independent Markovian noise into an equivalent time-independent Pauli noise model. Then, in the classical post-processing, one can apply this effective time-independent Pauli noise model to learn the noisy Pauli weights, ensuring robust shallow shadow predictions.


\end{proof}

\begin{lemma}[Twirling of time-dependent Markovian noise \cite{nonMarkovianRC}]\label{lemma:RC2}
    Assume time-dependent errors can be described with a two-parameter family of dynamical maps $\Phi(t_2,t_1)$, with property $\Phi(t+\tau,0)=\Phi(t+\tau,\tau)\Phi(\tau,0)$.
    Then, randomly sampling the twirling gates independently in each round tailors the time-dependent Markovian errors affecting each cycle into time-dependent stochastic Pauli noise when the errors on the easy gates are gate-independent.
\end{lemma}
This lemma is illustrated in \Cref{fig:effective_channel_twirling} (b), where each layer experiences time-dependent noise $\Lambda(t)$. After twirling, the effective noise model remains stochastic Pauli noise but with time-dependent probabilities. Moreover, there are no classical correlations between the probabilities across different layers.
\begin{theorem}[Noise-robust shallow shadow on time-dependent Markovian noise]\label{thm:time-noise}
In a noise-robust shallow shadow setting, a time-dependent stochastic Pauli noise model can be used in the data post-processing to effectively capture the effects of time-dependent non-Markovian noise that happened in the physical circuit. If the noise parameters are accurately learned, the predictions given by the robust shallow shadow remain unbiased.
\end{theorem}
\begin{proof}

We model the time-dependent noise as a dynamical map $\Phi(t_2,t_1)$, which has the property $\Phi(t_2+t_1,0)=\Phi(t_2+t_1,t_1)\Phi(t_1,0)$; this property also defines a type of non-Markovian noise \cite{nonMarkovianRC}. As in the proof of \cref{thm:nonmarkovian-noise_app}, we can insert single-qubit twirling gates after each single-qubit gate in a measurement circuit $U$. This insertion does not affect the ensemble average, so the value of the noisy Pauli weights remains unchanged. Using \cref{lemma:RC2}, we can effectively model this time-dependent noise using a Pauli noise model with time-dependent noise strengths for each measurement circuit. Then, in the classical post-processing, one can apply this effective time-dependent Pauli noise model to learn the noisy Pauli weights, ensuring robust shallow shadow predictions.
\end{proof}

\begin{lemma}[Twirling of non-Markovian noise]\label{lemma:RC3}
Consider a quantum channel where non-Markovian noise acts before and after a unitary operation $\mathcal{U}$. In general, this noise can be expressed in the Pauli basis as the map $\rho\rightarrow \sum_{i,j,k,l}\chi_{ijkl}P_j U(P_i \rho P_k)U^{\dagger}P_l$. After applying single-qubit Clifford twirling, as illustrated in \Cref{fig:effective_channel_twirling} (c), the effective noise model becomes a stochastic Pauli noise model with correlated probabilities across layers, given by $\rho\rightarrow \sum_{i,j}p_{i,j}P_j U(P_i \rho P_i)U^{\dagger}P_j$ where the joint Pauli noise probabilities are given by $p_{i,j}=\chi_{ijij}$.

    
\end{lemma}
This lemma follows directly from the application of \Cref{lemma:RC1} together with tensor diagram representations. A similar result has also been established in recent work \cite{nonMarkoviansuppresion}. We summarize the effective noise model in \Cref{fig:effective_channel_twirling} (c), highlighting that the primary distinction is the emergence of classical correlations in the effective Pauli noise probabilities across different layers for non-Markovian noise.

\begin{theorem}[Noise-robust shallow shadow on non-Markovian noise]\label{thm:non-markovian}
In a noise-robust shallow shadow setting, a classical correlated stochastic Pauli noise model can be used in the data post-processing to effectively capture the effects of non-Markovian noise that happened in the physical circuit. If the noise parameters are accurately learned, the predictions given by the robust shallow shadow remain unbiased.
\end{theorem}
\begin{proof}
    As in the proof of \cref{thm:nonmarkovian-noise_app}, we can insert single-qubit twirling gates after each single-qubit gate in a measurement circuit $U$. This insertion does not affect the ensemble average, so the value of the noisy Pauli weights remains unchanged. Using \cref{lemma:RC3}, we can effectively model this non-Markovian noise using a Pauli noise model with classical correlated noise strengths across different layers for each measurement circuit. Then, in the classical post-processing, one can apply this effective noise model to learn the noisy Pauli weights, ensuring robust shallow shadow predictions.
\end{proof}

\subsection{Bayesian Noise Learning}\label{app:bayes}
The first step of our Bayesian noise learning method is to collect a calibration dataset $\scD_{c}$ simply by running the shallow shadows protocol for a state $\rho=\ketbra{0}^{\otimes N}$, which we assume can be prepared with high fidelity. We then use this dataset to define a likelihood function $p(\scD_{c} \mid \lambda)$ for Bayesian inference as follows. We construct an estimator for the Pauli weights $\tilde{\omega}_{\scD_{c}}(P)$ by inverting \cref{eq:expec}:
\begin{equation}
    \tilde{\omega}_{\scD_{c}}(P) = \frac{1}{\abs{\scD_{c}}\Tr(\ketbra{0} P)} \sum_{i=1}^{\abs{\scD_{c}}}\Tr(\hat{\sigma_i} P),\label{eq:omega-cal}
\end{equation}
where we always choose $P$ to be some tensor product of identity and $Z$ operators. Since any Pauli weight depends only on the support of $P$ (thanks to the local-basis invariance of our ensemble) this subset of operators is sufficient to capture all Pauli weights. We calculate the standard deviations $\sigma_{P}$ of these estimators using a standard bootstrap estimate, by resampling (with replacement) the empirical data $1000$ times. The likelihood function is then simply
\begin{equation}
    p(\scD_{c} \mid \lambda) \propto \exp(-\sum_{P} \frac{\qty(\tilde{\omega}_{\scD_{c}}(P) - \omega_\lambda(P))^2}{2\sigma_{P}^2}),
\end{equation}
where the sum runs over all Paulis $P$. We design a tensor network algorithm that efficiently encodes $\omega_{\lambda}(P)$ for all Paulis $P$ given noise parameters $\lambda$ as a matrix product state (MPS), which enables the efficient calculation of likelihood function (details in \cref{app:tensor-network}). In practice, we choose all Pauli operators that are contiguous $Z$-strings up to weight 6 (beyond this weight, empirically recovered Pauli weights are too small to be resolved to within statistical significance). The prior $p(\lambda)$ is given by a log-normal distribution with $\sigma=2$, and centered around noise parameters $\tilde{\lambda}$ calculated via an independent inference process, namely that of Ref. \cite{berg2022probabilistic}. Together, the two fully specify the posterior distribution over the noise parameters:
\begin{equation}
    p(\lambda \mid \scD_{c}) \propto p(\scD_{c} \mid \lambda) p(\lambda).\label{eq:posterior}
\end{equation}
We can then sample from this posterior distribution using Hamiltonian Monte Carlo \cite{betancourt2018conceptual,phan2019composable}. Although this method essentially gives us access to the full posterior distribution, in practice, we typically pick a fixed value of $\lambda$ when inferring observables. This fixed value can be chosen a number of ways, including by taking the mean or median of the posterior distribution samples, or simply using a maximum a posteriori probability (MAP) estimate (found by maximizing \cref{eq:posterior} via gradient descent). We find that each of these three methods for fixing $\lambda$ produces almost identical results. 

\begin{figure}
    \centering
    \includegraphics[width=\linewidth]{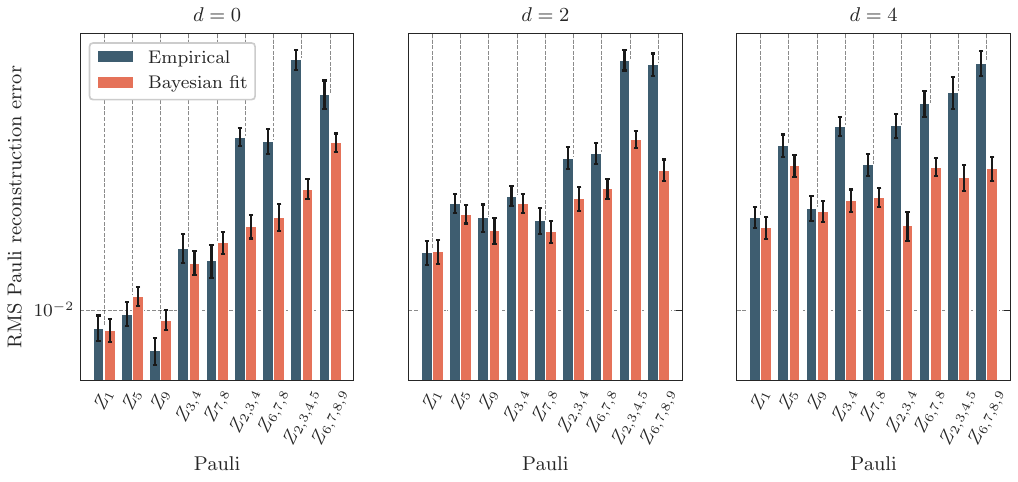}
    \caption{Performance improvement with Bayesian noise learning. We show the reconstruction error of various Pauli expectation values for $d \in \qty{0,2,4}$ using both the direct calibration procedure, and the Bayesian inference procedure, using the same number of measurements for each.}
    \label{fig:bayesian-appendix}
\end{figure}

\begin{figure}
    \centering
    \includegraphics[width=0.85\linewidth]{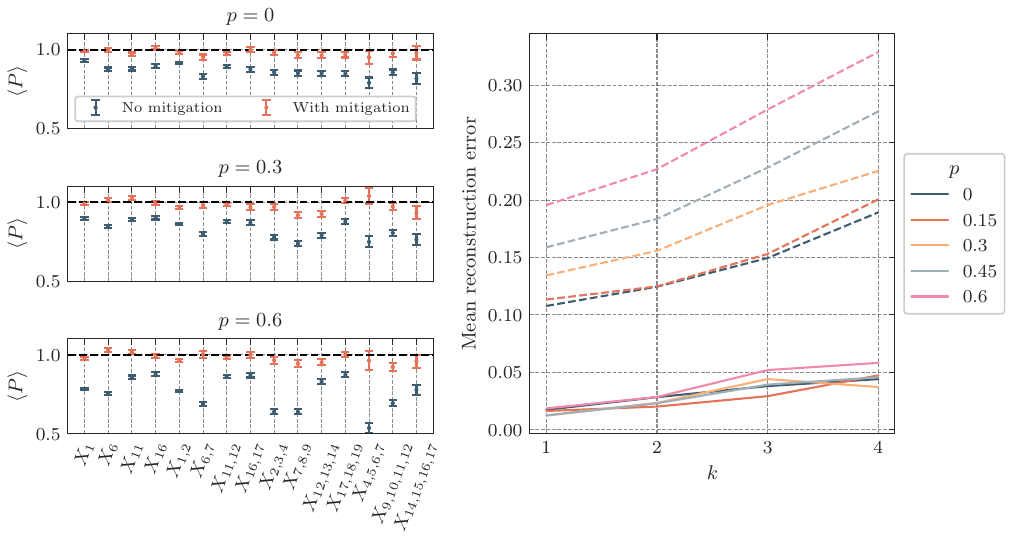}
    \caption{Expectation value of various Pauli $X$ operators for the $\ket{+}^{\otimes N}$ state. We simulate shadow circuits which have two-qubit Pauli noise with a typical noise scale $\sim 10^{-2.5}$. However, we add in a highly nonlocal, correlated noise source. At each layer of the circuit, with probability $p$, this noise source applies a random (three-qubit) Pauli operator on a random set of three qubits. Notably, we see that after calibration, our phenomenological Pauli-Lindblad noise model is able to account for the effects of this noise even at $p=0.6$, despite the fact that it assumes the noise is local. This can be seen on the right subfigure: the mean Pauli reconstruction error after mitigation (solid lines) is virtually independent of $p$, while the unmitigated results (dashed lines) grow quickly with $p$.}
    \label{fig:3-qubit}
\end{figure}

Although our Bayesian inference method makes an explicit assumption about the structure of the noise, in practice, this assumption turns out to be fairly weak. The reason is that the exact details of the noise channel are unimportant; only its effect on the Pauli weights $\omega_\lambda(P)$  are relevant. In \cref{fig:3-qubit}, we present evidence that $\omega_\lambda(P)$ depends only weakly on the detailed structure of the noise. To do this, we simulate a highly correlated and nonlocal noise source, which acts (with probability $p$) on random sets of three qubits with a random Pauli error at each layer of the circuit. Despite the fact that the sparse Pauli-Lindblad noise model cannot explicitly model this type of nonlocal noise, \cref{fig:3-qubit} shows that after using our Bayesian inference method to obtain a \emph{phenomenological} model for the noise, we are nevertheless able to recover unbiased estimates of various Pauli expectation values. In the future, it is also interesting to investigate the systematic error due to this model violation \cite{modelviolation}.

\section{Tensor network post-processing for robust shallow shadow\label{app:tensor-network}}
\subsection{Noise-free circuit Pauli weight calculation}
We begin with noise-free shallow circuit shadows, where both $\sC_{U,V}[\cdot]=\sC_{U,P}[\cdot]=U\cdot U^{\dagger}$. For a system comprising a single qubit, the Pauli weight is given by
\eqs{
\omega(P)=\underset{U}{\dsE}\Tr\left(\ket{0}\bra{0}^{\otimes 2}U^{\otimes 2}P^{\otimes 2}U^{\dagger\otimes 2}\right).
}

It is important to note that the tensor product of two replicas differs from that of different qubits. To distinguish between these cases, we use superscripts $(1)$ and $(2)$ to explicitly denote replicas and subscripts to denote qubit indices. For instance, $P_{(1,2)}^{\otimes 2}=X^{(1)}_1\otimes X^{(2)}_1\otimes Y^{(1)}_2\otimes Y^{(1)}_2$.

To understand the evolution, it is helpful to track the "distribution" of Pauli operators (induced by our probability distribution $P(U)$ over unitaries $U$) as we evolve forward under the random quantum circuit. For example, after applying one layer of single-qubit random Haar/Clifford twirling gates, a given Pauli operator, say $Z$, maps to an equal superposition of $X$, $Y$ and $Z$. Mathematically, this is expressed as
\eqs{
\mathbb{E}_{U\sim \text{Haar}_1}\left[U^{\otimes 2}Z^{\otimes 2}(U^{\dagger})^{\otimes 2}\right]=\dfrac{X^{(1)}\otimes X^{(2)}+Y^{(1)}\otimes Y^{(2)}+Z^{(1)}\otimes Z^{(2)}}{3}=-\dfrac{1}{3}I+\dfrac{2}{3}S,\label{eq:app_single_qubit}
}
where $I$ is the identity operator and $S$ is the swap operator between two copies.

More generally, for global Haar/Clifford distributions of $U$, we have
\eqs{
\mathbb{E}_{U\sim \text{Haar}_n}\left[U^{\otimes 2}P^{\otimes 2}(U^{\dagger})^{\otimes 2}\right]=\left(\dfrac{\Tr(P)-2^{-n} \Tr(PS)}{2^{2n}-1}\right)I+\left(\dfrac{\Tr(PS)-2^{-n}\Tr(P)}{2^{2n}-1}\right)S.\label{eq:app_multi_qubits}
}

For locally scrambled ensembles of $U$, which do not distinguish between different Pauli bases, the coefficients of $X$, $Y$, and $Z$ are equal. This makes it convenient to work in the $I$ and $S$ basis. We introduce a vector notation, $\ket{I}$ and $\ket{S}$, to denote the identity and swap basis (note that these are not superoperator notations). Using this notation, \cref{eq:app_single_qubit} becomes
\eqs{
\mathbb{E}_{U\sim \text{Haar}_1}\left[U^{\otimes 2}Z^{\otimes 2}(U^{\dagger})^{\otimes 2}\right]=-\dfrac{1}{3}\ket{I}+\dfrac{2}{3}\ket{S}=\begin{bmatrix}
    -\frac{1}{3}\\ \frac{2}{3}
\end{bmatrix}.
}

Applying \cref{eq:app_multi_qubits}, we can verify that two-qubit random Haar/Clifford unitaries act as follows:
\eqs{
\ket{II}&\rightarrow \ket{II}\\
\ket{SS}&\rightarrow\ket{SS}\\
\ket{IS}&\rightarrow \dfrac{2}{5}\left(\ket{II}+\ket{SS}\right)\\
\ket{SI}&\rightarrow \dfrac{2}{5}\left(\ket{II}+\ket{SS}\right),
}
which we can express in matrix form as
\eqs{
T_{12} = \begin{bmatrix}
    1 & \frac{2}{5} & \frac{2}{5} & 0\\
    0 & 0 & 0 & 0\\
    0 & 0 & 0 & 0\\
    0 & \frac{2}{5} & \frac{2}{5} & 1
\end{bmatrix}.
}

We call this action matrix $T$ the transfer matrix. Using transfer matrices simplifies tracking the distribution of our Pauli operator $P$ under random unitary gates. For shallow circuits comprising two-qubit random Haar unitaries with a brickwall structure, the transformation becomes a tensor network of nearest-neighbor transfer matrices $T_{i,i+1}$. The final state can be written as
\eqs{
\mathbb{E}_{U}\left[U^{\otimes 2}P^{\otimes 2}(U^{\dagger})^{\otimes 2}\right]\equiv\ket{\Phi}=c_1\ket{II\dots}+c_2\ket{IS\dots}+\cdots +c_{2^n}\ket{SS\dots}.
}

Since $\Tr\left(\ket{0}\bra{0}^{\otimes 2n}O\right)=1$ for any $O$ that is a tensor product of identities and swaps, the Pauli weight of $P$ simplifies to $\omega(P)=\sum_{i}c_i$. This can be computed by evaluating the inner product $\braket{+}{\Phi}$, where $\ket{+}\equiv\begin{bmatrix}
    1\\1
\end{bmatrix}^{\otimes n}$. We have implemented this algorithm using \texttt{PyClifford} \cite{pyclifford} and \texttt{Tensornetwork} \cite{tensornetwork}.

\subsection{Noisy circuit Pauli weight calculation}

When the two-qubit gates are not random Haar/Clifford, the distribution of Paulis after applying those gates will not be equally distributed in $X$, $Y$ and $Z$ direction. For example, the two-qubit gates in our experiments are CNOT gates twirled by single-qubit random Clifford gates. Unlike Haar random unitaries, the CNOT gates pick a preferred direction for Paulis, in the sense that they treat $X$ and $Z$ differently. Furthermore, our local correlated Lindbladian noise model treats different Pauli basis differently. So rather than working with identity $I$ and SWAP $S$ basis, we work with a four-state system, with basis states being the Paulis $\ket{I}$, $\ket{X}$, $\ket{Y}$, $\ket{Z}$. For instance, $\ket{Z}\equiv \begin{bmatrix}
    0 &0&0&1
\end{bmatrix}^{T}$. One should notice that the defined basis is different from superoperator basis. Then, the transfer matrix of a single-qubit twirling gate is
\eqs{
T=\begin{bmatrix}
    1 &0 &0 &0 \\
    0 & \frac{1}{3}& \frac{1}{3}& \frac{1}{3}\\
    0 & \frac{1}{3}& \frac{1}{3}& \frac{1}{3}\\
    0 & \frac{1}{3}& \frac{1}{3}& \frac{1}{3}\\
\end{bmatrix}.
}
The transfer matrix of a CNOT gate is a $16\times 16$ matrix defined by
\begin{equation}
    T_{\mathrm{CNOT}}[P_i,P_j] = \begin{cases}1 &\qq{if $P_j = \mathrm{CNOT} \cdot P_i \cdot \mathrm{CNOT}$ up to some phase} \\
    0 &\qq{otherwise.}
    \end{cases}
\end{equation}
Now, we also have a nice way of exactly solving Pauli distributions through noise channel, since the noise model is diagonal in the Pauli basis. For instance, if a qubit has single-qubit error rates $\lambda_X$, $\lambda_Y$, $\lambda_Z$, the transfer matrix of the noise channel for this site is
\eqs{
\Lambda = \diag(1,\exp(-2(\lambda_Y+\lambda_Z)),\exp(-2(\lambda_X+\lambda_Z),\exp(-2(\lambda_X+\lambda_Y))).
}
We can do the same for two-qubit noise, where our channel is then represented by a diagonal $16 \times 16$ matrix. At the end, our state $\ket{\Phi}$ will be a superposition over all Paulis. The only nonzero contributions to \eqref{eq:pauli-weight} are those where all the Pauli characters are either $I$ or $Z$. So, we can evaluate \eqref{eq:pauli-weight} with $\braket{+}{\Phi}$ where 
\begin{equation}
    \ket{+} = \mqty[1 \\0 \\0\\1]^{\otimes n}.
\end{equation}
This algorithm has also been implemented with \texttt{PyClifford} \cite{pyclifford} and \texttt{Tensornetwork} \cite{tensornetwork}.

\section{Optimal circuit depth of robust shallow shadows \label{app:optimal-depth}}
When predicting a Pauli observable $P_{\alpha}$ using classical shadows $\hat{\sigma}_{U,b}$, the inverse Pauli weight $\omega(P_{\alpha})$ acts as a rescaling factor:
\eqs{
\Tr(\rho P_{\alpha})=\dfrac{1}{\omega(P_\alpha)}\mathbb{E}_{\hat{\sigma}_{U,b}}\Tr(\hat{\sigma}_{U,b}P_{\alpha}).
}
This implies that $\text{Var}(\Tr(\rho P_{\alpha}))=\frac{1}{\omega(P_\alpha)^2}\text{Var}(\Tr(\hat{\sigma}_{U,b} P_{\alpha}))$, and the shadow norm is given by $\norm{P_\alpha}_{\text{shadow}}=\frac{1}{\omega(P_{\alpha})}$. Thus, calculating the Pauli weight $\omega(P_\alpha)$ directly determines the sample complexity for predicting the Pauli observable $P_\alpha$. We first review how to estimate $\omega(P_{\alpha})$ for brickwall shallow circuits in the noise-free case to determine the optimal circuit depth, then extend this analysis to include noise.

For a Pauli operator of size $k$, the shadow norm is given by 
\eqs{
\omega(P_k)&=\int dU \bra{0}UP_kU^{\dagger}\ket{0}^2=\int dU \bra{0}P_{k}(t)\ket{0}^2\\
&=\int dU \bra{0}\sum_{l}\alpha_{l,k}(t)P_l\ket{0}^2=\int dU\int d\Psi \bra{\Psi}\sum_{l} \alpha_{l,k}(t)P_l\ket{\Psi}^2\\
&=\sum_{l}\overline{|\alpha_{l,k}(t)|^2}\left(\frac{1}{3}\right)^l
=\mathbb{E}\left[\left(\dfrac{1}{3}\right)^{l}\right]\geq \left(\frac{1}{3}\right)^{\bar{l}},}
where $\overline{\cdots}=\int dU[\cdots]$ and $\bar{l}$ represents the average length of the evolved Pauli operator. Here, $\overline{|\alpha_{l,k}(t)|^2}$ represents the probability of finding a weight-$l$ Pauli operator in the evolution $P_k(t)=UP_k U^{\dagger}$. 

Each on-site non-trivial Pauli contributes a factor of $1/3$ to the Pauli weights. Since we do not distinguish between Pauli operators $X$, $Y$, and $Z$, we can represent all non-trivial Paulis as particles ($\darkcirc$) and the identity as holes ($\whitecirc$). A random two-qubit gate preserves the identity matrix but maps any of the $(2^4-1)$ non-trivial Pauli operators to an equal superposition of non-trivial Pauli operators: $T\ket{\darkcirc \whitecirc}=a\ket{\darkcirc \whitecirc}+a\ket{\whitecirc \darkcirc}+(1-2a)\ket{\darkcirc \darkcirc}$, where $a = (2^2-1)/(2^4-1)=1/5$. 

In the particle-hole basis ($\ket{\whitecirc \whitecirc},\ket{\darkcirc \whitecirc},\ket{\whitecirc \darkcirc},\ket{\darkcirc \darkcirc}$), the transfer matrix for a two-qubit Haar unitary is
\eqs{
T = \mqty[
    1 &0 & 0 &0\\
    0 &a &a &a\\
    0 &a &a &a\\
    0 &1-2a &1-2a &1-2a].
}

Using $T$, we can construct the transfer matrix $T_{\text{layer}}$ for one layer of even-odd two-qubit gates. The eigenstates and eigenvalues of $T_{\text{layer}}$ determine the relaxation of the bulk particle density. When applying $m$ layers to the all-particle state $T_{\text{layer}}^m\ket{\cdots \darkcirc \darkcirc \darkcirc \darkcirc \cdots}$, eigenstates with eigenvalues less than one decay and vanish. Thus, to understand bulk density relaxation, we need only consider the non-trivial eigenstate corresponding to the largest eigenvalue (which equals one in the noise-free case). The all-hole state $\ket{\cdots\whitecirc \whitecirc \whitecirc\whitecirc\cdots}$ is the trivial eigenstate, while the non-trivial eigenstate with eigenvalue one is $\ket{\psi}\propto 3\ket{\cdots\darkcirc\whitecirc\cdots}+3\ket{\cdots\whitecirc\darkcirc\cdots}+9\ket{\cdots\darkcirc\darkcirc\cdots}+\cdots$. Mean-field analysis shows that at equilibrium, the probability of finding a particle at any site is $3/4$. The relaxation rate $\gamma$ is determined by the second-largest eigenvalue of $T_{\text{layer}}$. The bulk relaxation can then be expressed as 
\eqs{
n(t)=\dfrac{3}{4}+\dfrac{1}{4}e^{-\gamma t},
}
where $n(t)$ represents the average particle density at time $t$.

\subsection{Exact upper bound}
For the regime $k\gg t\gg 1$, \citet{Ippoliti_2023} mapped the transfer matrix model to a classical random walk model and showed that
\eqs{
n(t) = \dfrac{3}{4}+t^{-3/2}e^{-\gamma t},~e^{-\gamma}\equiv \left(\dfrac{4}{5}\right)^2.
}
Consequently, for an initial state with $k$ particles, the average particle length after depth $t$ is
\eqs{
\mathbb{E}[l]=\left(\dfrac{3}{4}+t^{-3/2}e^{-\gamma t}\right)(k+2v_b t),
}
where $v_b$ characterizes the linear spreading of the operator size. 

For simplicity, we model noise as a local depolarizing channel with parameter $\lambda$. This introduces an additional damping factor $e^{-\lambda}$ that penalizes each non-identity Pauli (particle) during the stochastic evolution:
\begin{equation}
    \omega_t(P_k)=\mathbb{E}\left[\left(\dfrac{1}{3}\right)^{l_t}\exp\left(-\lambda \sum_{i=1}^{t}l_i\right)\right], \label{eq:omegat}
\end{equation}
where $l_t$ denotes the weight of the evolved Pauli operator at time step $t$. 

We can establish a lower bound on the Pauli weight by noting that $\sum_{i=1}^{t}l_i\leq \sum_{i=1}^{t}k+2i\leq kt+t^2$, since the Pauli observable's support grows by at most 2 per layer of the brickwall circuit. Applying this bound and Jensen's inequality yields
\eqs{
\omega_t(P_k)\geq \exp (-\lambda (kt+t^2))\left(\dfrac{1}{3}\right)^{\mathbb{E}[l]}.
}

In summary, 
\begin{subequations}
    \begin{align}
        \omega_t(P_k) &\geq \exp(-\lambda(kt+t^2) - \log(3) \qty[\qty(\frac{3}{4} + t^{-3/2} \qty(\frac{4}{5})^{2t})\qty(k+2v_B t)]) \\
        &\geq \exp(-(k+t) \qty[\lambda t + \log(3)\qty(\frac{3}{4} + t^{-3/2} \qty(\frac{4}{5})^{2t})])
    \end{align}
\end{subequations}

\subsection{Phenomenological model}
The bound obtained using Jensen's inequality proves overly pessimistic for \cref{eq:omegat}. We can achieve tighter bounds by rewriting 
\begin{equation}
    \omega_t(P_k) = \mathbb{E}\qty[\exp(-l_t \log 3 - \lambda \sum_{i=1}^t l_i)].
\end{equation}
We define $X(t) = -l_t \log 3 - \lambda \sum_{i=1}^t l_i$, and model this as a normal distribution with mean $\mu_k(t)$ and variance $\sigma^2_k(t)$. Then, $\omega_t$ is the expectation of a log-normal distribution, which can be calculated in closed form: $\omega_t(P_k) = \mathbb{E}[\exp(X)]= \exp(\mu_k(t) + \sigma_k^2(t)/2)$; we note that Jensen's inequality corresponds to the bound $\mathbb{E}[\exp(X)] \geq \exp(\mu_k(t))$, so we can get significant improvements in our bounds by appropriately accounting for $\sigma_k^2(t)$. Estimating $\omega_t$ reduces to modeling $\mu_k(t)$ and $\sigma_k(t)$ phenomenologically. We start with $\mu_k(t)$: $\mathbb{E}[X] = -\log 3 \cdot \mathbb{E}[l_t] - \lambda \sum_{i=1}^t \mathbb{E}[l_i]$. The expectation of the particle length $l_t$ is modeled extremely well by the `relaxation' and `spreading' effects described in \cite{Ippoliti_2023}:
\begin{equation}
    \mathbb{E}[l_t] = \underbrace{(k+v_b t)}_{\text{Spreading}}\underbrace{\qty(\frac{3}{4} + \frac{1}{4} \exp(-t/t_0))}_{\text{Relaxation}},
\end{equation}
where $k$ is the initial operator size, $v_b$ is some `butterfly velocity', and $t_0$ is some typical timescale for the relaxation of the occupation towards the equilibrium value $\frac{3}{4}$. Therefore,
\begin{subequations}
    \begin{align}
\mu_k(t) &= - \log 3 \cdot \frac{k+v_b t}{4} \qty(3 + \exp(-t/t_0)) - \frac{\lambda}{4} \sum_{i=1}^t \qty(k+v_b i)\qty(3+\exp(-i/t_0)) \\
    &\geq - \log 3 \cdot \frac{k+v_b t}{4} \qty(3 + \exp(-t/t_0)) - \frac{3\lambda}{4}\qty(kt + v_b t \cdot (t+1)/2) - \frac{\lambda}{4} \int_0^\infty (k+v_b t') \exp(-t'/t_0) \dd{t'} \\
    &= - \log 3 \cdot \frac{k+v_b t}{4} \qty(3 + \exp(-t/t_0)) - \frac{\lambda}{4} \qty(3\qty(kt + v_b t \cdot (t+1)/2) + v_b t_0^2 + kt_0) \\
    &\geq - \log 3 \cdot \frac{k+v_b t}{4} \qty(3 + \exp(-t/t_0)) - 3\lambda t \qty(kt + v_b (t+1)/2) \qq{(since $t_0 = 0.65$)}\\
    &\geq -\qty(k+v_b t) \qty(\frac{\log 3}{4} (3+\exp(-t/t_0)) + 3\lambda t).
    \end{align}
\end{subequations}

The variance $\sigma_k^2(t)$, in theory, depends on both $l_t$ as well as the sum $\sum_{i=1}^t l_i$, and covariances between the two. However, since we typically assume $\lambda \sim 10^{-2}$ while $\log 3 \sim 1$, we ignore the contribution of $\lambda$. Furthermore, since $l_t$ and $\sum_{i=1}^t l_i$ have a positive correlation, ignoring the contribution from $\sum_i l_i$ can only result in an underestimate of the variance, hence an underestimate of $\omega_t(P_k)$ -- since we are interested only in lower bounding $\omega_t(P_k)$, we can do this freely. We find empirically that the variance of $l_t$ is well-described by an initial fast period of growth that saturates at some value which depends linearly on $k$, and is then followed by linear growth. That is,
\begin{equation}
    \text{var}(l_t) = (mk+b)(1-\exp(-t)) + V' t.
\end{equation}
\begin{figure}
    \centering
    \includegraphics[width=0.9\linewidth]{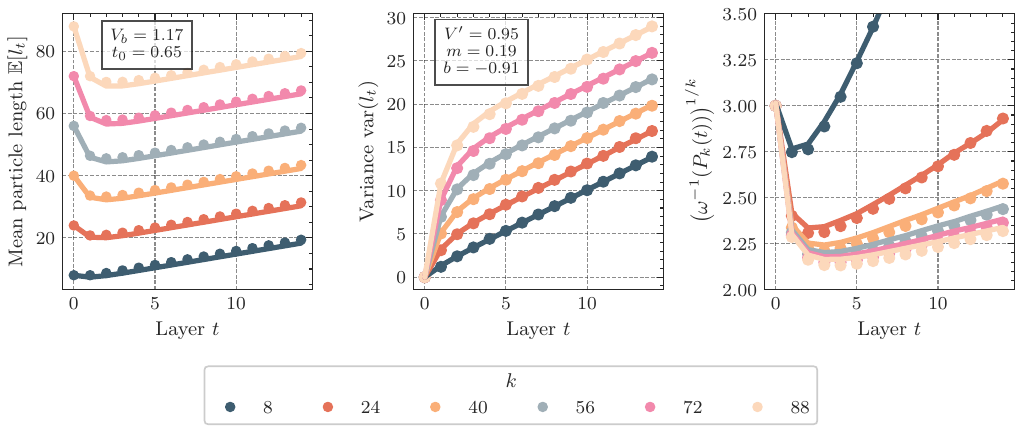}
    \caption{Phenomenology for the random walk model in \cref{eq:omegat}. We model distribution of particle length $l_t$ at a given time step as a normal distribution. The shadow norm $\omega(P_k(t))$ is then the expectation of a log-normal distribution, which can be calculated exactly; theoretical predictions based on this model are shown in solid lines, while the circle markers indicate simulated values obtained via Monte Carlo simulation of the random walk.}
    \label{fig:phenom}
\end{figure}

Putting this together, we have
\begin{equation}
    \omega_t(P_k) \geq \exp(-\qty(k+v_b t) \qty(\frac{\log 3}{4} (3+\exp(-t/t_0)) + 3\lambda t) + \frac{(\log 3)^2 (mk+b)(1-\exp(-t)) + V't}{2}).
\end{equation}
These results are compiled in \cref{fig:phenom}.

\begin{figure}
    \centering
    \includegraphics[width=0.4\linewidth]{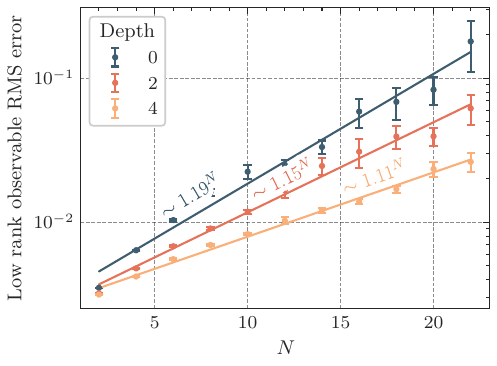}
    \caption{Fidelity estimation error with $d \in \qty{0,2,4}$. Note that although fidelity estimation costs grow exponentially in $N$, the base of this exponential dependence improves significantly with larger depth, enabling one to achieve practically efficient estimates by choosing a depth that is appropriate to the system size of interest.}
    \label{fig:low-rank}
\end{figure}

Finally, in \cref{fig:low-rank}, we show how an increasing circuit depth results in strict improvements for nonlocal property estimation, such as fidelity. 

\section{Experimental platform}\label{app:experiment}
\begin{figure}
    \centering
\subfloat{\includegraphics[width=.45\textwidth]{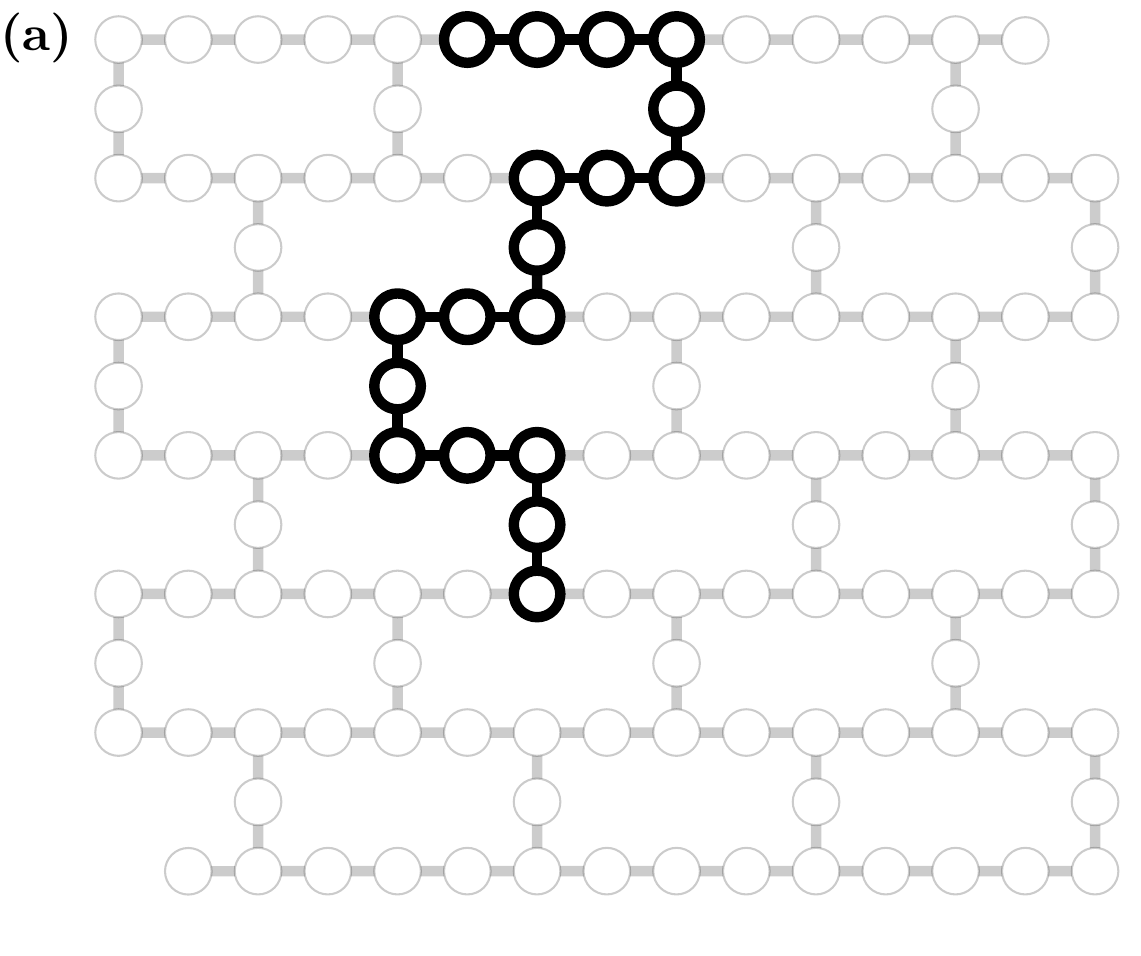}}
    \hspace{2em}
    \subfloat{\includegraphics[width=.365\textwidth]{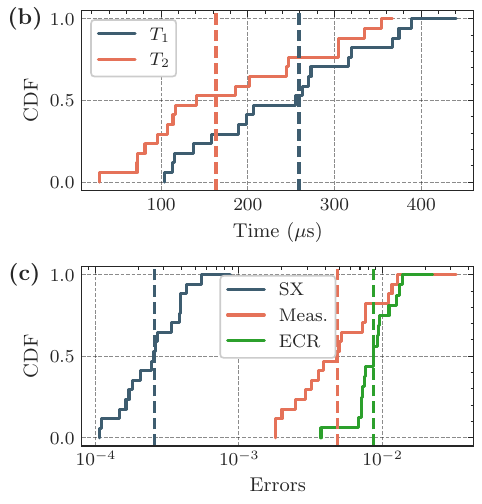}}
    \caption{Experimental implementation details. In (a), we show the device layout of \textit{ibm\_kyiv} where 127 qubits are organized in a heavy-hex architecture. The chosen 18 qubit string is highlighted. In (b), we plot the cumulative distribution of $T_1$ and $T_2$ coherence times of the chosen qubits and in (c) we plot the single qubit gate (SX), two qubit entangling gate (ECR) and readout error rates.}
    \label{fig:device}
\end{figure}

We perform all experiments on \textit{ibm\_kyiv}~\cite{ibmquantum}, a 127-qubit superconducting quantum processor. We choose a chain of 18 qubit whose properties are shown in \cref{fig:device}. All the circuits are decomposed into the device's native gateset composed of the two-qubit entangling Echoed-Cross Resonance (ECR) gate, $1/\sqrt{2}(IX-XY)$ that is locally equivalent to controlled-NOT, and single qubit gates $\sqrt{X}$ (SX) and arbitrary rotations in the $Z$ basis. All the ECR gates are performed in 561.778 nanoseconds. The speed of device as characterized by
Circuit Layer Operations Per Second (CLOPS) is 5000, which crucially enables fast sampling over many different circuits~\cite{wack2021quality}. For each application circuit and fixed depth of Cliffords, we applied 10000 different random unitaries and measured 100 shots each. It took us, on average, 6 minutes of quantum hardware usage time to collect these 1 million samples.  The experiments corresponding to $\ket{+}^{\otimes 18}$ and $\ket{\phi}$ in \cref{fig:recovered,fig:sample-complexity} were performed on December 21\textsuperscript{st}, 2023 and those for the AKLT state correponding to \cref{fig:entanglement} were performed on February 6\textsuperscript{th}, 2024. 

\begin{figure}
    \centering
    \includegraphics[width=\linewidth]{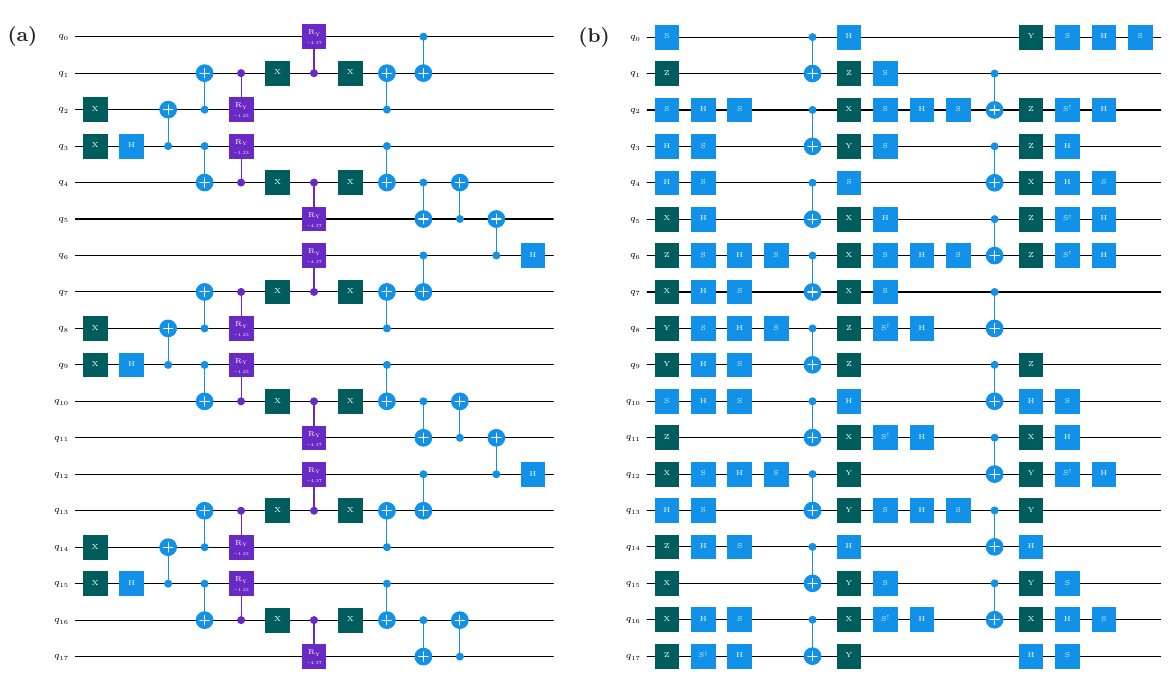}
    \caption{In (a), we show the AKLT resource state preparation circuit. In (b), we show one instance of a circuit sampled from our shallow circuit ensemble ($d=2$).}
    \label{fig:circuits}
\end{figure}
\end{document}